\documentclass{article}
\usepackage[utf8]{inputenc}

\usepackage{authblk}
\usepackage{caption}
\usepackage{subcaption}
\usepackage{graphicx}
\usepackage{placeins}
\usepackage{multirow}
\usepackage{booktabs}
\usepackage{tabu}
\usepackage{hyperref}
\usepackage{url}
\usepackage{makecell}
\usepackage[table,xcdraw]{xcolor}

\title{Family based HLA imputation and optimization of haplo-identical transplants
}

\author[1]{\fontsize{10pt}{10pt}\textbf{Zuriya Ansbacher-Feldman}}
\author[1]{\textbf{Sapir Israeli}}
\author[2, 3]{\textbf{Martin Maiers}}
\author[2, 3, 4]{\textbf{Loren Gragert}}
\author[5]{\textbf{Dianne De Santis}}
\author[6, 7]{\textbf{Moshe Israeli}}
\author[1]{\textbf{Yoram Louzoun}}

\affil[1]{Department of Mathematics, Bar-Ilan University, Ramat Gan, Israel}
\affil[2]{Center for Blood and Marrow Transplant Research, Minneapolis, MN, US}
\affil[3]{National Marrow Donor Program/Be The Match, Minneapolis, MN, USA}
\affil[4]{Department of Pathology and Laboratory Medicine, Tulane Cancer Center, Tulane University School of Medicine, New Orleans, LA, USA}
\affil[5]{Department of Clinical Immunology, PathWest, Fiona Stanley Hospital, Perth, Australia}
\affil[6]{Tissue Typing Laboratory, Beilinson Hospital, Rabin Medical Center, Petach-Tikva, Israel}
\affil[7]{Department of Digital Medical Technologies, Holon Institute of Technology, Holon, Israel}

\begin{document}

\maketitle
\begin{abstract}
Recently, haplo-identical transplantation with multiple HLA mismatches has become a viable option for system cell transplants. Haplotype sharing detection requires imputation of donor and recipient. We show that even in high-resolution typing when all alleles are known, there is a 15\% error rate in haplotype phasing, and even more in low resolution typings. 
Similarly, in related donors, parents haplotypes should be imputed to determine what haplotype each child inherited. We propose GRAMM (GRaph bAsed FaMilly iMputation) to phase alleles in family pedigree HLA typing data, and in mother-cord blood unit pairs.  
We show that GRAMM has practically no phasing errors when pedigree data are available. We apply GRAMM to simulations with different typing resolutions as well as paired cord-mother typings, and show very high phasing accuracy, and improved alleles imputation accuracy. 
We use GRAMM to detect recombination events and show that the rate of falsely detected recombination events (False Positive Rate) in simulations is very low. We then apply recombination detection to typed families to estimate the recombination rate in Israeli and Australian population datasets. The estimated recombination rate has an upper bound of 10-20\% per family (1-4 \% per individual).

GRAMM is available at: \url{https://gramm.math.biu.ac.il/}. 

\end{abstract}

\section{Introduction}
HLA matching is crucial for both Hematopoietic Stemm Cell (HSCT) and organ tranplants\cite{kamoun2017hla,manski2019predicting,kekre2016impact,nowak2008role}. Recently, full haplotype matching has also been shown to be important for transplant\cite{lorentino2017impact}. Targeted HLA typing does not contain haplotype phase information, as intergenic regions are not typed by current methods. As such, haplotype phasing of the HLA genes have to be imputed \cite{maiers2019grimm}. Moreover, HLA typing methods often do not unambiguously determine the alleles present, either due to incomplete sequencing or inability to phase within the gene\cite{madbouly2014validation}. 
While the accuracy of imputation algorithms for high resolution HLA allele determination has been previously tested \cite{alter2017hla, maiers2019grimm}, the HLA haplotype phase determination accuracy has not been tested. We here analyze the haplotype phasing accuracy of donor HLA imputation, and then propose a family based imputation algorithm to improve this accuracy when family pedigree data are available for related donors transplants.

Family based imputation tools have been previously developed both in the context of HLA and more generally for genome-wide genotyping data\cite{sargolzaei2014new,ullah2019comparison,saad2014combining,gorjanc2017prospects,liu2019revisit}. Computer based analysis of multilocus HLA pedigree data was first developed in 1984 for the Ninth International Histocompatibility Workshop.  The utility of a computer based approach is not only to scale up analysis beyond what can be practically done manually, but the ability to generate probabilistic outputs when all or part of the input HLA genotyping is missing or ambiguous \cite{neugebauer1984analysis}.

The Bayesian SNP (single nucleotide polymorphisms) phasing algorithm PHASE \cite{Phase} reconstructs SNPs haplotypes from genotypes using population reference frequency data caters to binary SNPs and not HLA alleles.
More recently, HaplObserve \cite{osoegawa2019tools} builds HLA gene haplotypes from genotypes of nuclear families; however, HaplObserve requires information on at least two parents and a child, and only outputs the most likely haplotypes rather than a probability distribution or all possible parent haplotype pairs. Finally, recombination events, while detectable in manual pedigree analysis of HLA in families \cite{gao2009total}, has not so far been included in any automated HLA pedigree analysis software. As such, an advanced family imputation algorithm for phase and genotype in ambigous family typing is still required. Such a tool will facilitate and improve the accuracy of related donor imputation, and will allow for more precise estimates of haplotypes for haplotype frequency estimates. 

Even today, HLA typing is often ambiguous \cite{paunic2016charting}, and family members require HLA imputation in related donors. A simple solution could be to impute each family member by himself, and take the most probable haplotype pair in each. We show here that this leads to inconsistencies in an important fraction of cases. As such, phase and allele ambiguity have to be resolved simultaneously. A solution to that is to impute the parents based on their own typing and the typing of their descendants. This can be performed using the recently developed GRIMM imputation \cite{maiers2019grimm}. 

GRIMM is based on first opening all the possible phasing of the haplotypes and then determining for each phase independently a probability distribution of high resolution alleles to characterize ambiguity within each respective phase. We here propose to enlarge GRIMM to lock some of the phases of the parents using information on their descendants, and as such limit the possible phases of the parents. Then imputation can be performed on the legitimate phases. We denote the resulting family imputation algorithm   GRAMM (GRaph bAsed faMily iMputation). 

In the following section, we show the limitations of existing algorithms, and then explain the logic of GRAMM. We then propose multiple measures to compare the accuracy of GRAMM. Finally, we show that GRAMM can be used to estimate the recombination rate in realistic families.

\section{Results}
\subsection{Errors in phasing in imputation}
In order to test the accuracy of phase imputation in current algorithms, we computed the phase distribution in simulated families with GRIMM. Although the simulations contain families, in GRIMM, each individual was analyzed separately. 

The simulated HLA typings for parents were generated by sampling two high resolution 5-locus (A,C,B,DRB1,DQB1) haplotypes from NMDP White European full registry haplotype distribution. We then simulated one or more children within families by randomly choosing a pair of genotypes as parents. We then introduced typing ambiguity for the family as the typing would appear utilizing different HLA typing methods with varying levels of typing ambiguity \cite{israeli2021hla}. Finally, in some simulations, we also simulated recombination events. The simulations we used are detailed in Table~\ref{table: Simulation details}. We checked an additional set of simulations, with a higher number of children, and the results were similar (Supp. Mat. Table S1, Fig S1).

\begin{table}[ht]
\begin{tabular}{cllccc}
\hline
\begin{tabular}[c]{@{}c@{}}Simulation\\ ID\end{tabular} &
  \multicolumn{1}{c}{\begin{tabular}[c]{@{}c@{}}Simulation \\ name\end{tabular}} &
  \multicolumn{1}{c}{\begin{tabular}[c]{@{}c@{}}Typing \\ resolution\end{tabular}} &
  \begin{tabular}[c]{@{}c@{}}Parents\\ number\end{tabular} &
  \begin{tabular}[c]{@{}c@{}}Recombination\\ rate\end{tabular} &
  \begin{tabular}[c]{@{}c@{}}Children\\ number\end{tabular} \\ \hline
1  & TR-H\_P\_RR0\_C3     & High   & 2   & 0  & 3   \\ \hline
2  & TR-H\_P\_RR5\_C3     & High   & 2   & 5  & 3   \\ \hline
3  & TR-M\_P\_RR0\_C3     & Medium & 2   & 0  & 3   \\ \hline
4  & TR-M\_NP\_RR0\_C3    & Medium & 0   & 0  & 3   \\ \hline
5  & TR-M\_NP\_RR5\_C3    & Medium & 0   & 5  & 3   \\ \hline
6  & TR-L\_PP\_RR0\_C1-3  & Low    & 0-2 & 0  & 1-3 \\ \hline
7  & TR-L\_PP\_RR1\_C1-3  & Low    & 0-2 & 1  & 1-3 \\ \hline
8  & TR-L\_PP\_RR2\_C1-3  & Low    & 0-2 & 2  & 1-3 \\ \hline
9  & TR-L\_PP\_RR5\_C1-3  & Low    & 0-2 & 5  & 1-3 \\ \hline
10 & TR-L\_PP\_RR10\_C1-3 & Low   & 0-2 & 10 & 1-3 \\ \hline
\end{tabular}
\caption{\textbf{Simulations details.} Simulations are notated by 4 fields: (1) \textbf{TR}- typing resolution (High (H) = no ambiguity + fully typing (A,B,C,DRB1,DQB1); Medium (M)= ambiguity + fully typing; Low (L)= ambiguity + some missing C/DQB1), (2) \textbf{P/NP/PP}- Number of parents (P =  2 parents; NP =  no parents; PP = partial parents (0-2)), (3) \textbf{RR}- recombination rate (0/1/2/5/10 \%), (4) \textbf{C}- children number (3/1-3).}
\label{table: Simulation details}
\end{table}

\FloatBarrier

We first imputed the phase of each family member individually using GRIMM. Each simulation contains 500 families, with known genotypes and haplotypes of each family member (see Methods). 

\begin{figure}
    \centering
    \includegraphics[width=0.7\linewidth]{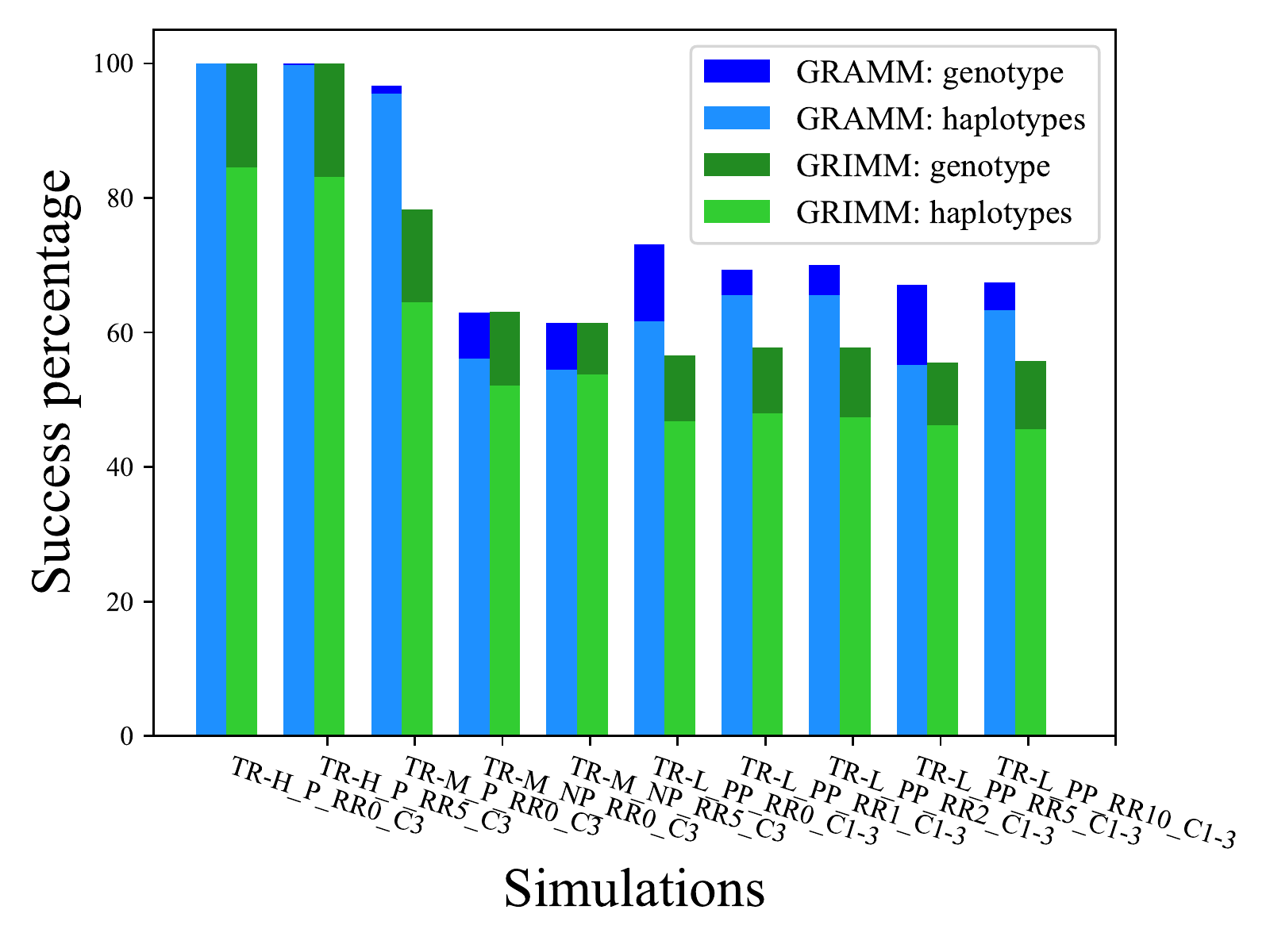}
    \caption{\textbf{Comparison between the accuracy, measured as the fraction of properly imputed individuals haplotypes/genotypes of GRAMM (blue) and GRIMM (green).} The numeric scores are present in Table~\ref{table: Success percentage of GRAMM and GRIMM + p value}. Deep blue and green represent accuracy in the unphased genotypes, and light blue green are for accuracy in the phased genotypes. The difference between deep and light are errors in phase, but not in alleles.}
    \label{fig: GRAMM VS GRIMM Success}
\end{figure}

We checked how often GRIMM returns the proper unphased high resolution allele-level genotype (Figure~\ref{fig: GRAMM VS GRIMM Success}, deep green bars) and how often GRIMM returns the proper haplotypes (light green in Figure~\ref{fig: GRAMM VS GRIMM Success}) as the most probable result. The score presented for each simulation is the fraction (in percentage) of families members with proper GRIMM based predictions. The real haplotypes are known, since those are simulated families.

\FloatBarrier

\begin{table}[ht]
\begin{tabular}{lcccccc}
\hline
\multicolumn{1}{c}{\multirow{2}{*}{\textbf{Simulation names}}} &
  \multicolumn{2}{c}{\textbf{Genotype}} &
  \multicolumn{1}{l}{\multirow{2}{*}{\textbf{P-value}}} &
  \multicolumn{2}{c}{\textbf{Haplotype}} &
  \multirow{2}{*}{\textbf{P-value}} \\
\multicolumn{1}{c}{} &
  \multicolumn{1}{l}{\textbf{GRAMM}} &
  \multicolumn{1}{l}{\textbf{GRIMM}} &
  \multicolumn{1}{l}{} &
  \multicolumn{1}{l}{\textbf{GRAMM}} &
  \multicolumn{1}{l}{\textbf{GRIMM}} &
   \\ \hline
TR-H\_P\_RR0\_C3     & 100\%  & 100\%  & 1.0     & 100\%  & 84.5\% & \textbf{9.4e-93}  \\ \hline
TR-H\_P\_RR5\_C3     & 100\%  & 100\%  & 1.0     & 99.7\% & 83.1\% & \textbf{1.8e-72}  \\ \hline
TR-M\_P\_RR0\_C3     & 96.7\% & 78.3\% & \textbf{3.8e-78} & 95.5\% & 64.5\% &\textbf{ 1.7e-148} \\ \hline
TR-M\_NP\_RR0\_C3    & 62.9\% & 63.1\% & 0.95    & 56.1\% & 52.1\% & 0.12     \\ \hline
TR-M\_NP\_RR5\_C3    & 61.4\% & 61.4\% & 1.0     & 54.4\% & 53.8\% & 0.86     \\ \hline
TR-L\_PP\_RR0\_C1-3  & 73.1\% & 56.6\% & \textbf{1.2e-17} & 61.6\% & 46.8\% & \textbf{2.7e-13}  \\ \hline
TR-L\_PP\_RR1\_C1-3  & 69.3\% & 57.8\% & \textbf{1.3e-08} & 65.6\% & 48\%   & \textbf{1.7e-17}  \\ \hline
TR-L\_PP\_RR2\_C1-3  & 70\%   & 57.8\% & \textbf{2.7e-09} & 65.6\% & 47.4\% & \textbf{4.7e-18}  \\ \hline
TR-L\_PP\_RR5\_C1-3  & 67.1\% & 55.5\% & \textbf{1.1e-08} & 55.2\% & 46.2\% & \textbf{1.9e-05}  \\ \hline
TR-L\_PP\_RR10\_C1-3 & 67.4\% & 55.8\% & \textbf{1.7e-07} & 63.3\% & 45.6\% & \textbf{7e-15}    \\ \hline
\end{tabular}
\caption{\textbf{Accuracy of GRAMM and GRIMM and chi-square p value of the comparison between them.} ``Genotype" columns is the allelic inference. ``Haplotype" is phasing accuracy}
\label{table: Success percentage of GRAMM and GRIMM + p value}
\end{table}

\FloatBarrier

In the unambiguous typing simulations where all alleles are known, the unphased genotype is properly detected in 100 \% of cases by GRIMM. However, even in those cases, 15-20 \% of individuals are not properly phased by GRIMM. In simulations with typing uncertainty and partial typing (i.e., not all alleles are typed, and those that are typed are in low resolutions), there is a high uncertainty as to which high resolution alleles are present (deep green). This results in the true high resolution genotype often being different from the most probable high resolution genotype. Beyond that, further uncertainty remains in the haplotype phasing of the alleles, even when the most probable set of high resolution alleles was the true set.

\begin{figure}[htp]
\centering

    \begin{subfigure}{0.45\columnwidth}
    \centering
    \includegraphics[width=\textwidth]{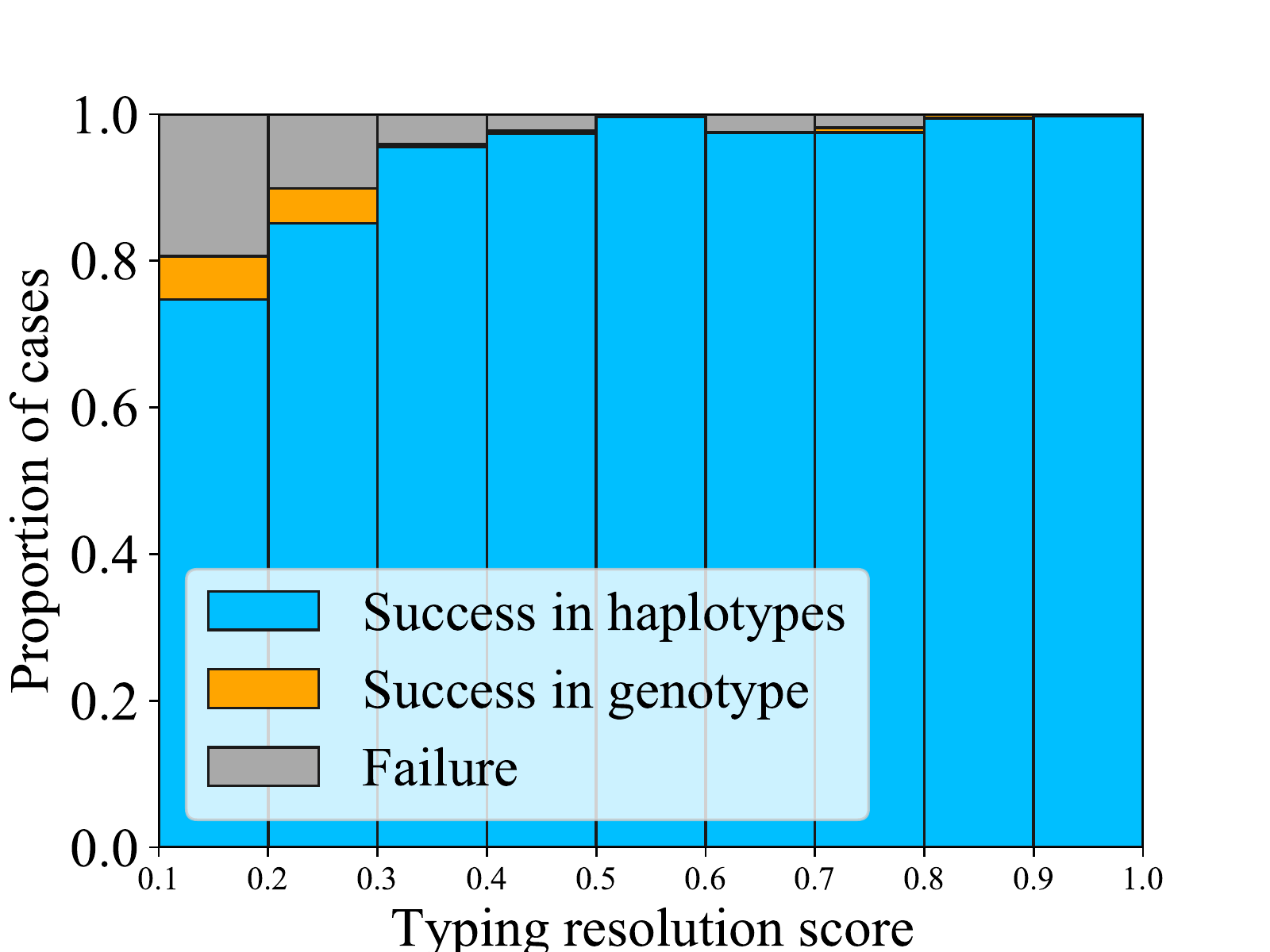}
    \caption{Simulation 3: TR-M\_P\_RR0\_C3}
    \label{fig:time1}
    \end{subfigure}
    \begin{subfigure}{0.45\columnwidth}
    \centering
    \includegraphics[width=\textwidth]{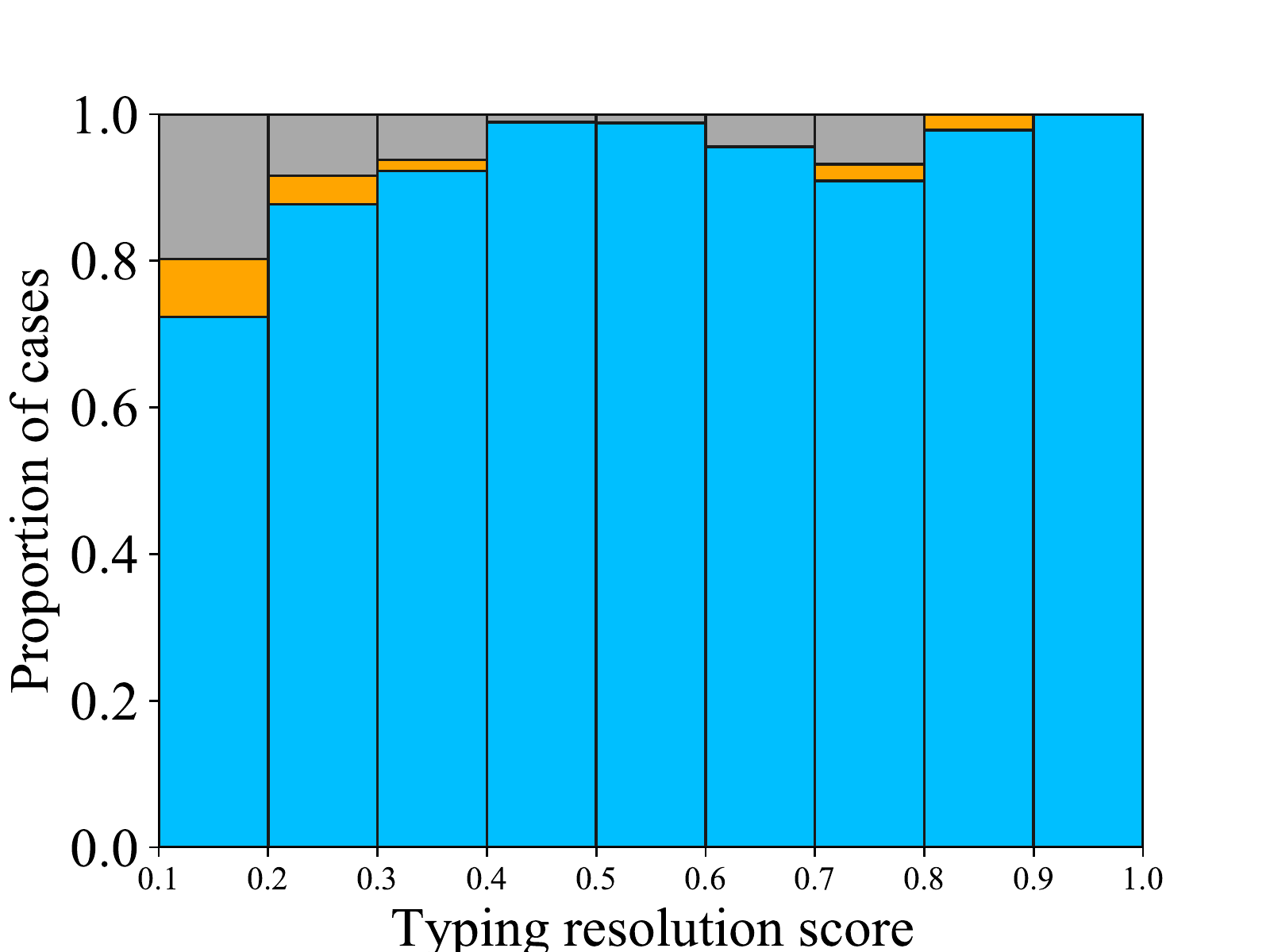}
    \caption{Simulation 4: TR-M\_NP\_RR0\_C3}
    \label{fig:time2}
    \end{subfigure}

    \begin{subfigure}{0.45\columnwidth}
    \centering
    \includegraphics[width=\textwidth]{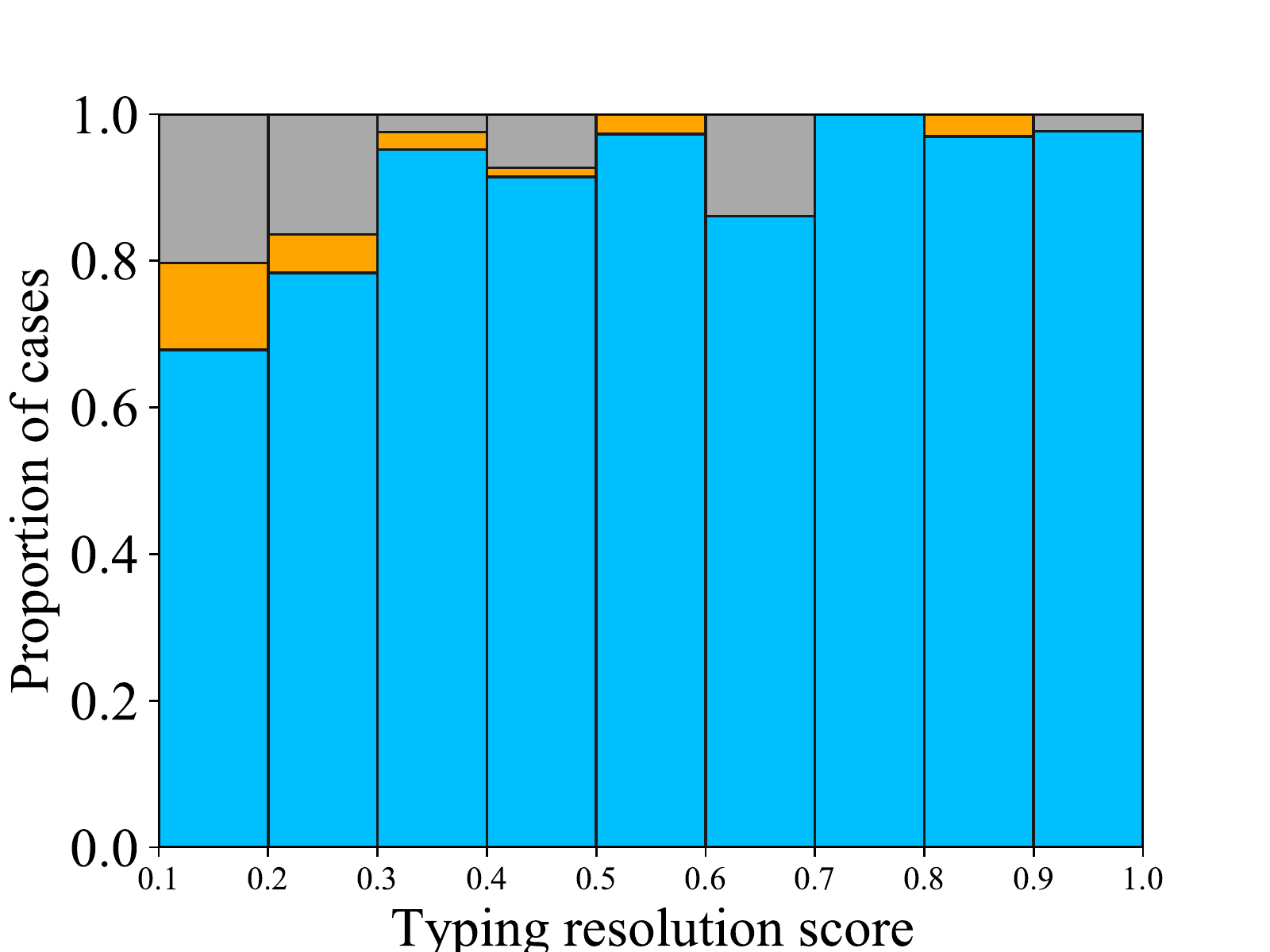}
    \caption{Simulation 5: TR-M\_NP\_RR5\_C3}
    \label{fig:time1}
    \end{subfigure}
    \begin{subfigure}{0.45\columnwidth}
    \centering
    \includegraphics[width=\textwidth]{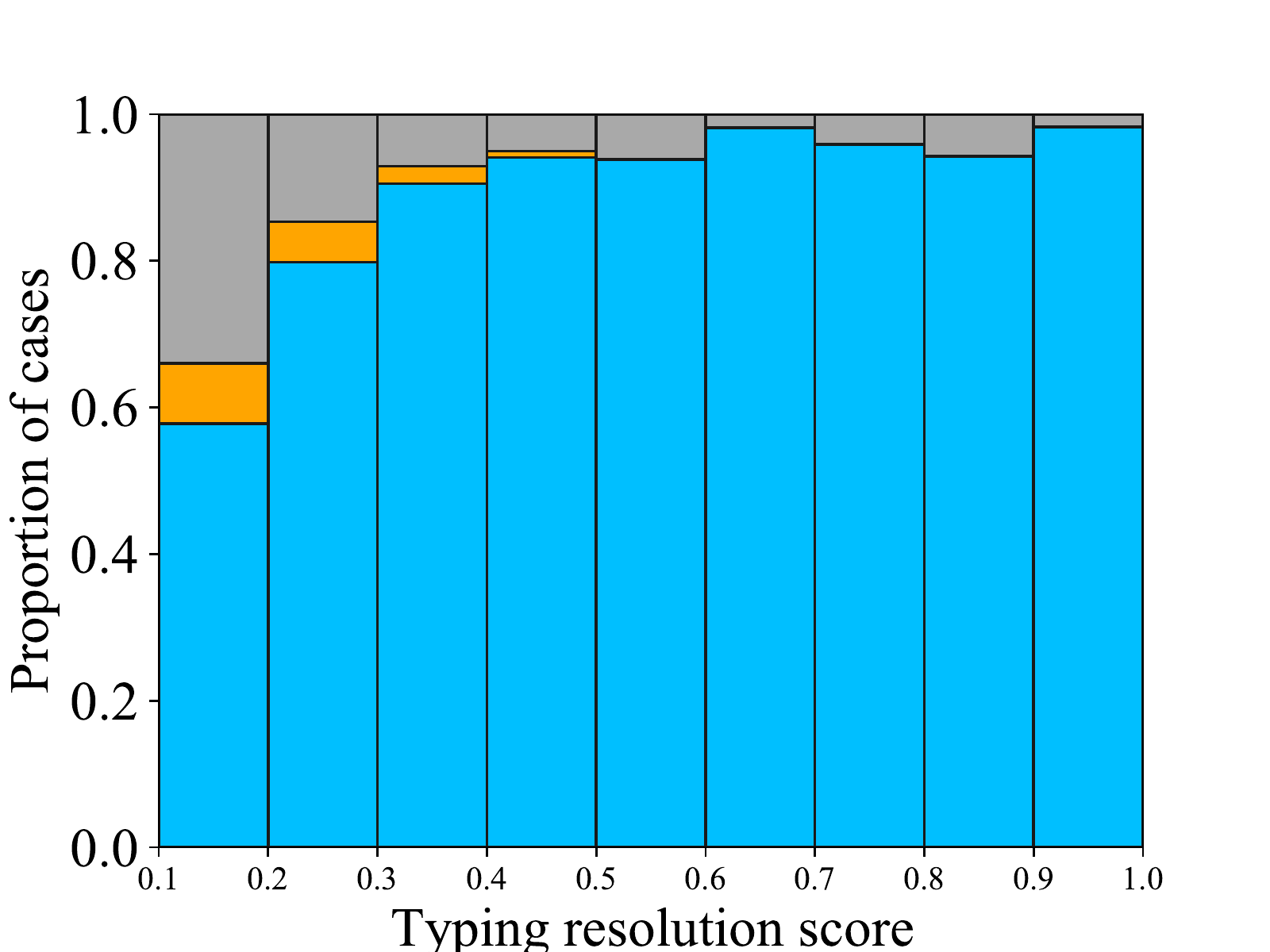}
    \caption{Simulation 9: TR-L\_PP\_RR5\_C1-3}
    \label{fig:time2}
    \end{subfigure}
    
    \caption{\textbf{Relationship between typing resolution and accuracy in haplotype phasing and high resolution alleles imputation in simulated datasets}. We computed for each simulated donor its typing resolution $TRS=\sum_i P_i^2$. For each set of donors grouped by $TRS$, we computed, using GRIMM, the fraction of properly computed phased multi-locus high resolution genotypes (Blue), the fraction of properly computed unphased multilocus high resolution genotype, but wrong haplotype phasing (Orange), and one or more wrong high resolution alleles (Gray). The error rate decreases with $TRS$, but increases in the more complex simulations.}
    \label{fig: Pi_Square}

\end{figure}

To understand if the high uncertainty in allele assignment results from rare haplotype that may be sub-sampled in the haplotype frequencies or from frequent haplotypes that may have multiple possible phasing of the same allele combinations in the frequencies, we computed for each simulated donor the normalized probability of each multilocus high resolution genotype $p_i$ (e.g., if a donor had 2 genotypes, the first with probability 0.02 and the second with probability 0.03, we computed the normalized probabilities to be 0.4 and 0.6). We then computed for each donor the typing resolution score \cite{paunic2016charting}, a metric representing the ambiguity of the typing, defined to be $TRS=\sum_i p_i^2$. The TRS score is 1 if there is a single possible multilocus high resolution genotype, and low if there are many candidates of differing probability.  We grouped donors, based on their value of $TRS$, and computed the fraction of errors as a function of $TRS$ (Figure~\ref{fig: Pi_Square}). Four simulations are shown here; the other simulations are in Supp. Mat., Fig. S2, S3). One can clearly see that the both allele and in phasing errors decrease with increasing $TRS$. In the simulations with recombinations and low typing resolution, allele identification and  phasing errors can happen even for high $TRS$ values (i.e., there is a dominant solution, but it is wrong).

\begin{figure}[htp]
\centering

    \begin{subfigure}{0.45\columnwidth}
    \centering
    \includegraphics[width=\textwidth]{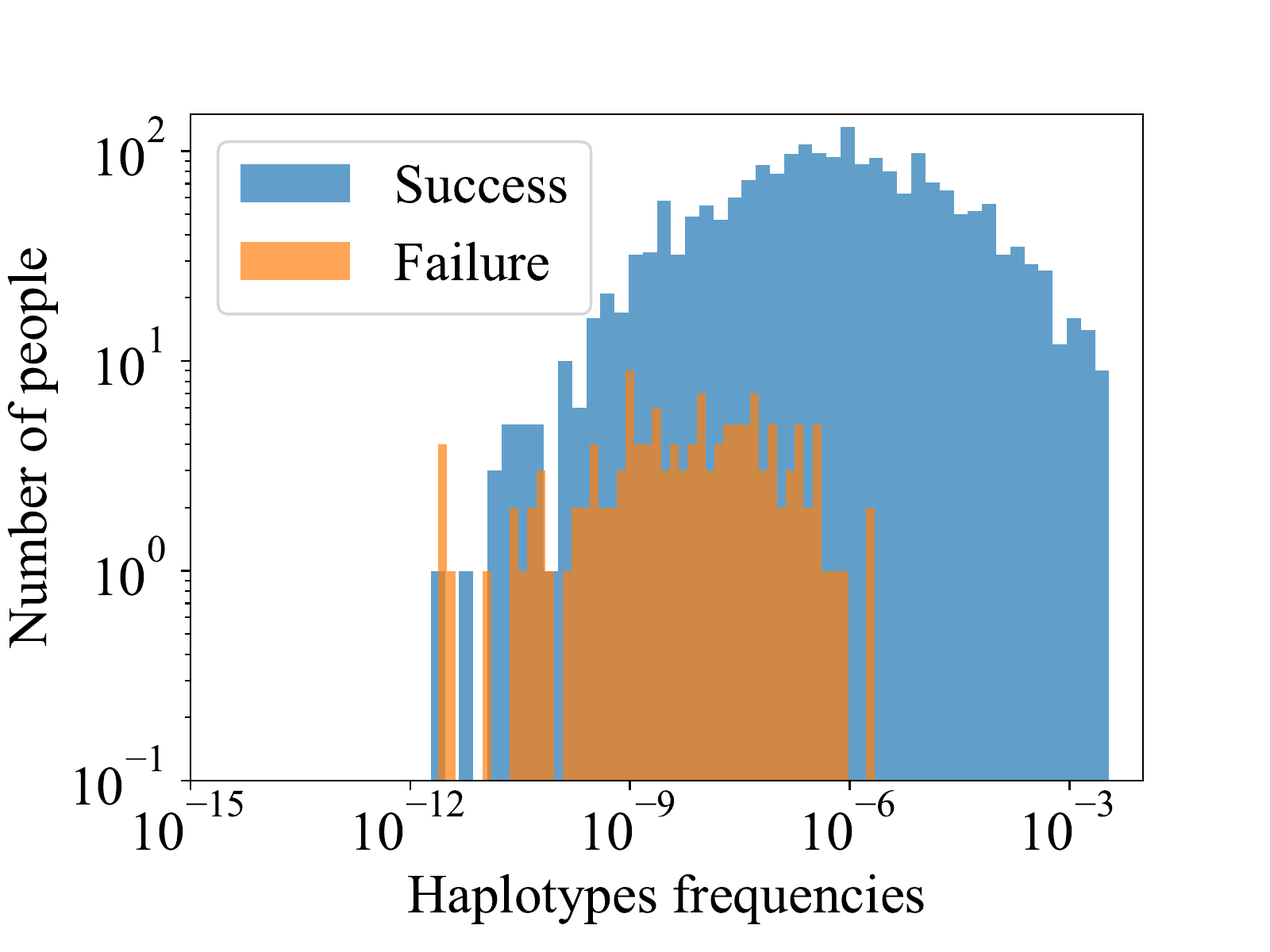}
    \caption{Simulation 3: TR-M\_P\_RR0\_C3}
    \label{fig:time1}
    \end{subfigure}
    \begin{subfigure}{0.45\columnwidth}
    \centering
    \includegraphics[width=\textwidth]{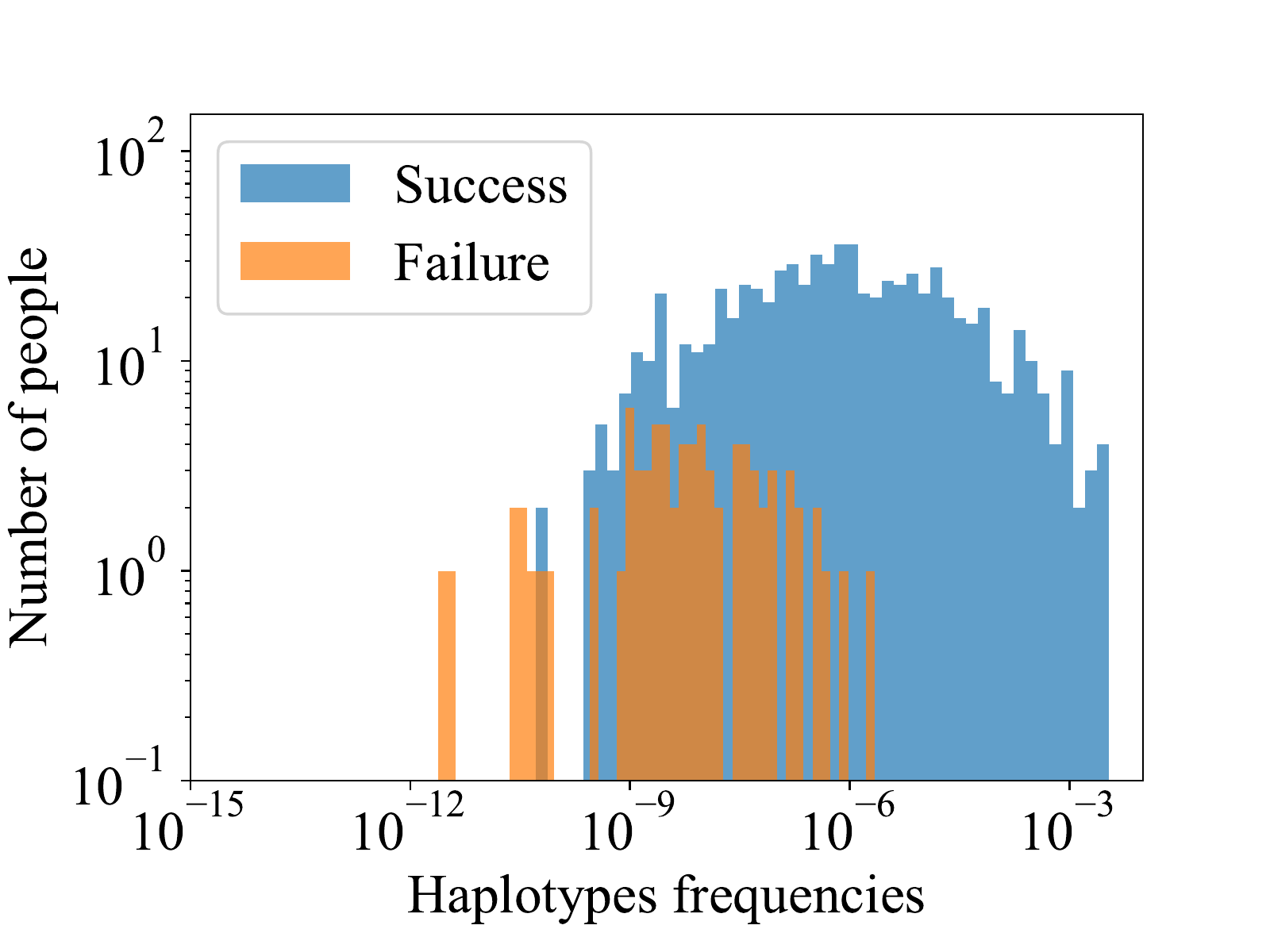}
    \caption{Simulation 4: TR-M\_NP\_RR0\_C3}
    \label{fig:time2}
    \end{subfigure}

    \begin{subfigure}{0.45\columnwidth}
    \centering
    \includegraphics[width=\textwidth]{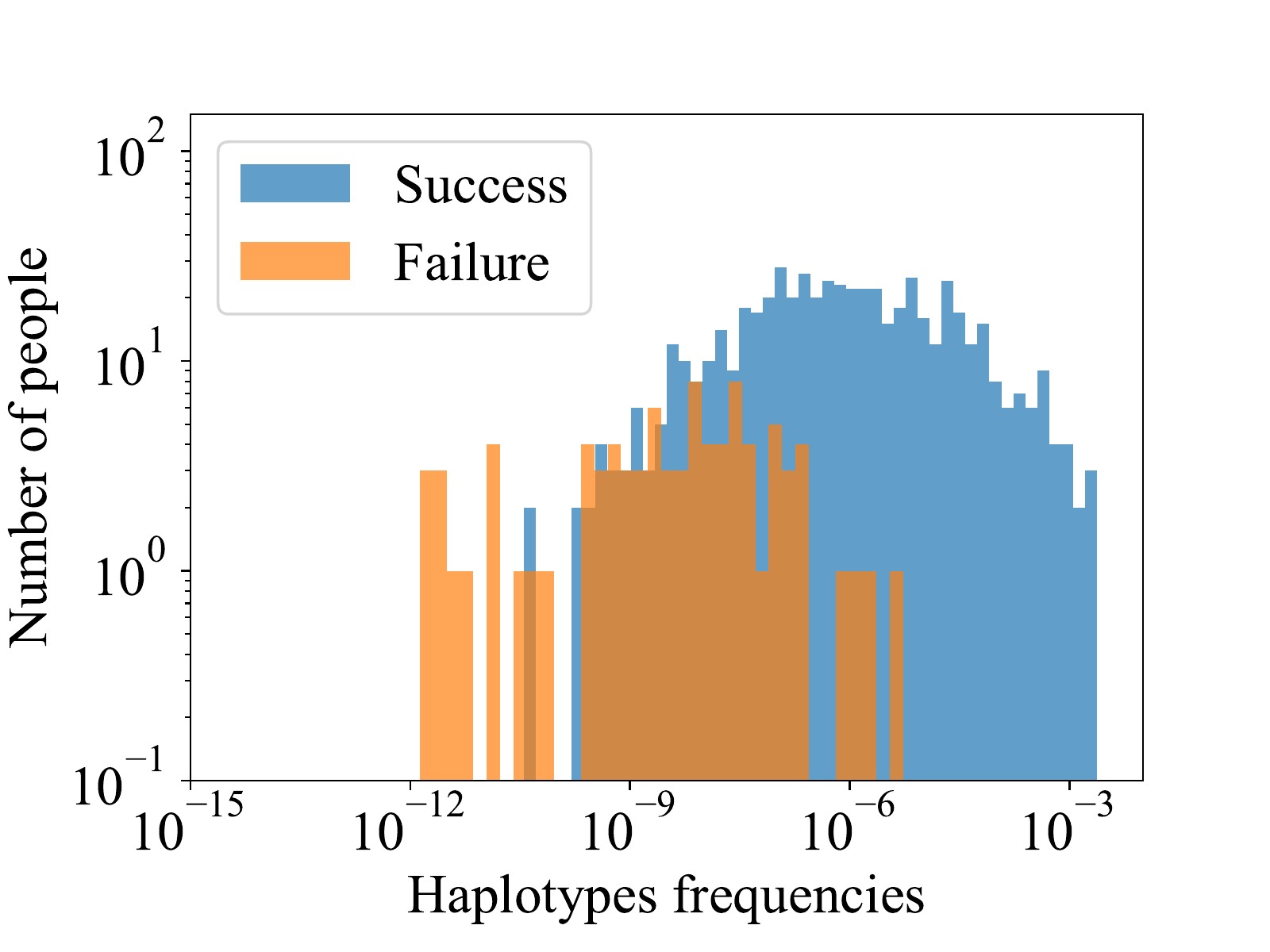}
    \caption{Simulation 5: TR-M\_NP\_RR5\_C3}
    \label{fig:time1}
    \end{subfigure}
    \begin{subfigure}{0.45\columnwidth}
    \centering
    \includegraphics[width=\textwidth]{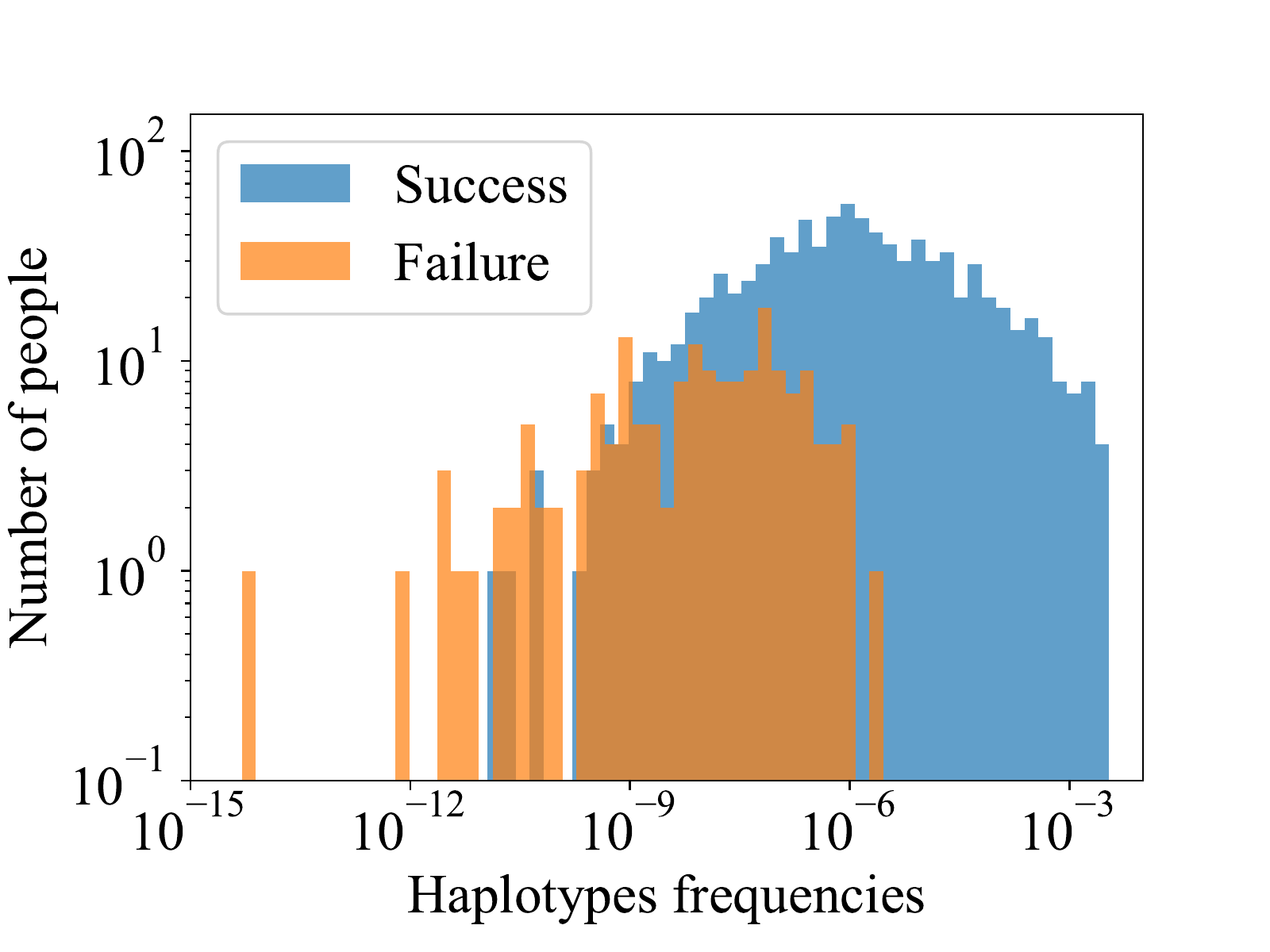}
    \caption{Simulation 9: TR-L\_PP\_RR5\_C1-3}
    \label{fig:time2}
    \end{subfigure}
    
    \caption{\textbf{Haplotype frequencies of correctly imputed genotypes (Blue bars) and erroneously imputed genotypes (Orange bars) in simulations.} The haplotypes were imputed using GRIMM. For each genotype, we computed the frequency of the two haplotypes composing the genotype. In homozygotes, the haplotype was counted twice in the histogram.}
    \label{fig: Frequencies VS Success}

\end{figure}

To further identify the sources of errors, we computed for each simulated donor, the phased genotype frequency assuming Hardy Weinberg Equilibrium (HWE) based on the true simulated high resolution haplotype pair. The allele and phasing errors are centered around rare genotypes (Orange vs Blue bars in Figure~\ref{fig: Frequencies VS Success}), but as the simulations get more complex (lower typing resolution and less alleles typed), errors are observed for more frequent genotypes.

\subsection{GRAMM resolves the phasing errors and reduces allelic errors.}
To address the errors in phase assignment when analyzing one donor at a time, we tested whether including family pedigree HLA information can improve phase and genotype accuracy. Assuming that recombination events are rare, family members can be assumed to share full high resolution HLA haplotypes. As such, when imputing all members of a family, one can assume phased HLA haplotype inheritance between parents and children. 

We propose a novel algorithm to simultaneously solve phase and allelic ambiguities in families, named  GRAMM (GRaph bAsed faMily iMputation). The basic logic of GRAMM is that given HLA typing of information of the children and parents within a nuclear family (parents and their direct biological children), it is enough to reconstruct the 4 haplotypes of the two parents ($F1$, $F2$, $M1$ and $M2$), and then resolve for each child which haplotypes were inherited (e.g., $F1$ and $M2$). To impute the parents' haplotypes, we use the GRIMM framework that first opens all phases of each parent and then solves the possible ambiguity in each phase conditional on the haplotype frequencies in the population. 

However, in the presence of information on family members, GRAMM can use the information from the children  to remove phases that are not consistent with the typing of the children. We will further denote that "phase locking". Given a family, GRAMM goes over family members one by one, the parents (if they exist) first, and then the children, and concludes which phases in the parents (even in the absence of any typing in the parents) are locked (in other words: which alleles must be in the same haplotype - not all the alleles must be locked in a haplotype). 

\begin{figure}[h!]
    \centering
      \includegraphics[scale=0.5]{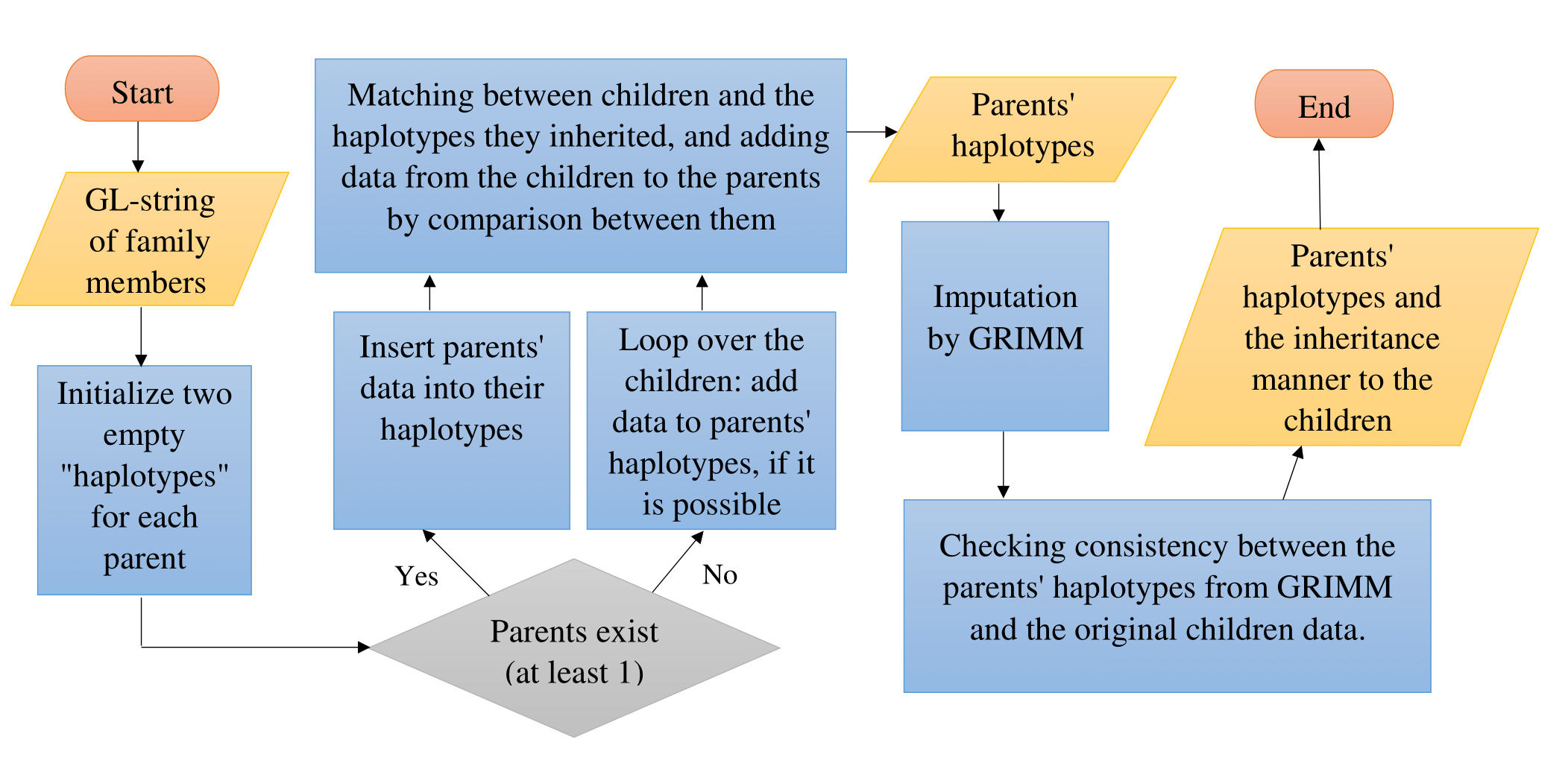}
      \caption{\textbf{GRAMM Algorithm}. The input is a set of HLA typings of family members represented in genotype list string (GL String) format. The imputation process consists of 5 stages: (1) Initialization: creation of  2 empty `haplotypes' for each parent, that include empty fields for all alleles: (currently A, B, C, DRB1, DQB1). (2)(a) If the input contains typing data for the parents, the single-locus allele (or lists of possible alleles if there is typing ambiguity) is inserted into the empty haplotypes. (b) else (no parents in data): combine information from children data and insert to parent haplotypes. (3) Loop over the children: associate each child with the haplotypes inherited from parents and add appropriate information on additional alleles to the haplotypes by comparison between the child and the parent data. (4) Insert the resulting candidate haplotype pairs of each parent to GRIMM, for imputation. (5) Check the results from GRIMM and reject imputation results on the parents that are inconsistent with at least one family member.}
      \label{fig: GRAMM Algorithm}
\end{figure}

GRAMM will then further detect family members whose HLA is not consistent with the others, and marks those as recombination events or allelic inconsistencies. The algorithm details are given in Methods, and in Figure~\ref{fig: GRAMM Algorithm}. 

We analyzed the allelic and phase imputation accuracy in GRAMM. In this case, each family was analyzed as a unit, but for consistency we report the accuracy counting each  family members once (i.e., a family with 4 members will have twice the weight of a 2 member family in the accuracy computation). The evaluation of imputation accuracy is done as in the GRIMM simulation analysis described previously: a comparison between the most probable result of the parents' haplotypes (from which can be derived the haplotypes of the whole family) and the true simulated high resolution haplotype pair. Haplotypes of a family member that are identical between the true haplotypes used in the simulations and the most probable prediction, are considered as a success (see Figure~\ref{fig: GRAMM VS GRIMM Success} and Table~\ref{table: Success percentage of GRAMM and GRIMM + p value}). 

When compared to individual imputation, introducing the family information removes completely the phase and allelic ambiguity in the easy simulations (Figure~\ref{fig: GRAMM VS GRIMM Success}, Deep Blue). Moreover, in the more complex simulation, the family information also removes some of the allelic ambiguity (Deep Blue vs Deep Green). Finally, beyond the allelic ambiguity, there is practically no phase ambiguity, even in very complex simulations (Deep vs Light Blue). Thus, at least in the related transplant setting with at least 3 or 4 family members (parents or children), the phasing can be typically performed accurately with two-generational pedigree data regardless of HLA typing ambiguity.

\subsection{Mother-Cord phasing is improved by GRAMM}
To test the applicability of GRAMM to real-life data and not only to simulations, we analyzed paired HLA-typed samples of mothers and their cord blood units. Typically, the cord blood was typed in high resolution and the mother was typed at low resolution. We analyzed 16,404 cord blood unit-mother pairs \cite{magalon2015banking} by GRIMM and GRAMM (individual vs family imputation). Then, we checked the consistency between the haplotypes of the mother and the child that GRIMM and GRAMM predicted (i.e., whether the most probable haplotype pairs of both the mother and the cord contained a shared haplotype). GRIMM produced 15,518 consistent families (94.5\%), while the accuracy of GRAMM was 16,189 consistent families (98.6\%) (Chi square, p=9.5e-94).

\subsection{Estimate of recombination rate}

\begin{figure}
    \centering
    \includegraphics[width=0.5\linewidth]{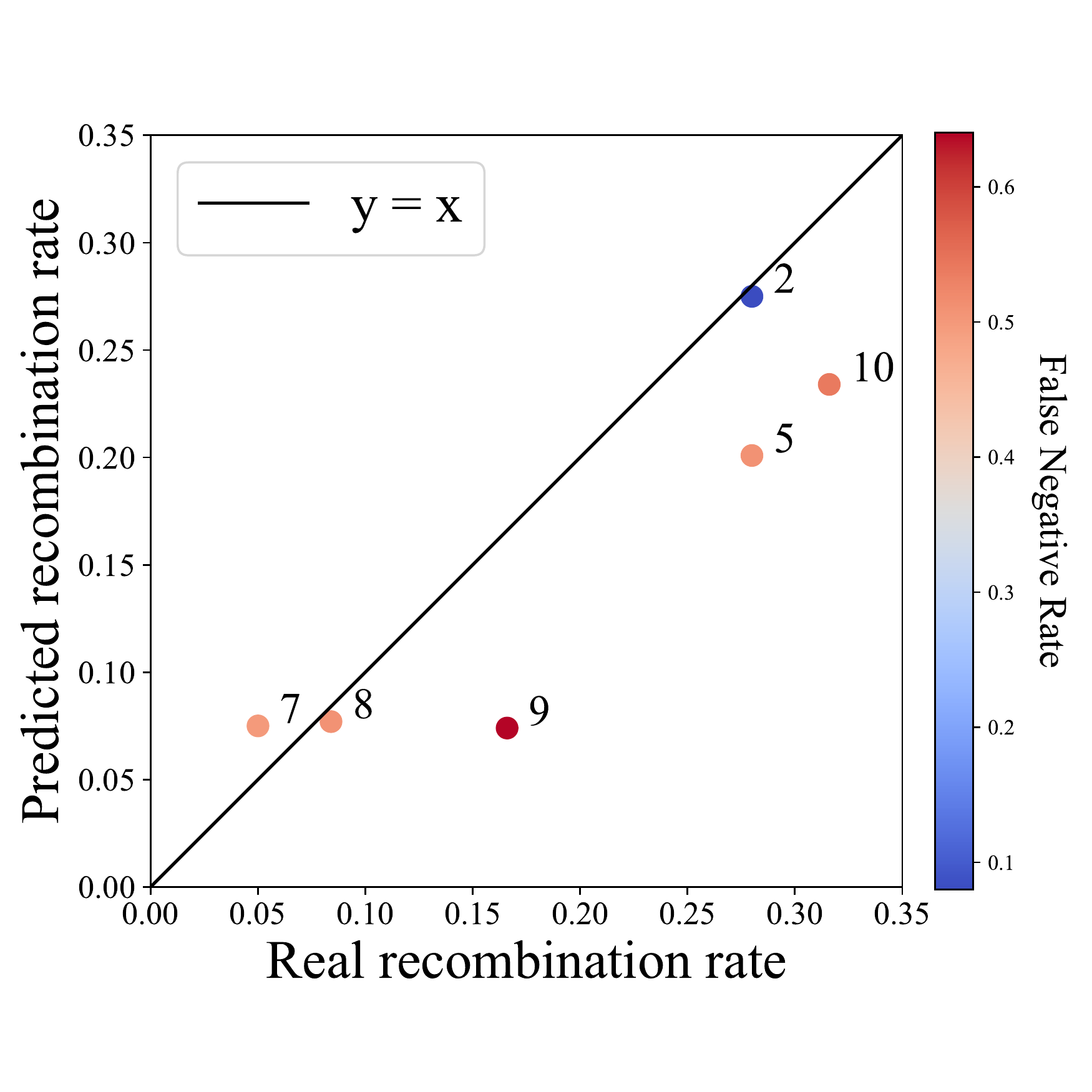}
    \caption{\textbf{Real vs predicted recombination rate.} Each point is a simulation (only those with recombination rate $>$ 0 : simulations 2, 5, 7, 8, 9, 10). Recombination rates are determined through families that contain at least one individual that cannot be explained without  a recombination event. The recombination rates here are per family and therefore are higher than those in Table \ref{table: Simulation details} that represents the individuals recombination rate.  The color describes the False Negative Rate - presented in Table \ref{table: FPR and FNR of GRAMM}. }
    \label{fig: Real VS predicted recombination rate}
\end{figure}

GRAMM can detect recombination events. If a family could only be explained assuming a single recombination event, a family member that resolves the recombination is removed from the analysis. If two or more recombination events are necessary, the entire family was disqualified. To test whether GRAMM properly recognizes recombination in families, we used the simulations that contain simulated recombination events (see Methods section~\ref{simulations explanation}). We computed for each simulation the fraction of families without recombination where GRAMM detected a recombination (The False Positive Rate $(FPR = \frac{False\ Positive}{False\ Positive + True\ Negative})$), and the fraction of families where there was a recombination and GRAMM missed it (the False Negative Rate $(FNR = \frac{False\ Negative}{False\ Negative + True\ Positive})$). GRAMM has practically no false positives. However, it has many false negatives when the typing resolution is not high (Table 2), since recombination events can be missed (for example, if the allele where a recombination occurred was ambiguously typed or not typed). Still, even in the most complex simulation, around 60\% of recombination events were detected, suggesting that GRAMM can be used to estimate the recombination rate in families (up to an under-estimate of 40\%. The under-estimate may be the result of partial or low resolution typing e.g. the allele were recombination occurred was not typed) (Figure ~\ref{fig: Real VS predicted recombination rate}).

\begin{table}[h!]
\centering
\begin{tabular}{lcc}
\hline
\textbf{} & \textbf{False Positive Rate} & \textbf{False Negative Rate} \\ \hline
\textbf{2: TR-H\_P\_RR5\_C3}      & 0    & 0.08 \\ \hline
\textbf{5: TR-M\_NP\_RR5\_C3}  & 0    & 0.51 \\ \hline
\textbf{7: TR-L\_PP\_RR1\_C1-3} & 0.04 & 0.5 \\ \hline
\textbf{8: TR-L\_PP\_RR2\_C1-3} & 0.03 & 0.51 \\ \hline
\textbf{9: TR-L\_PP\_RR5\_C1-3} & 0.006 & 0.64 \\ \hline
\textbf{10: TR-L\_PP\_RR10\_C1-3} & 0.05 & 0.54 \\ \hline
\end{tabular}
\caption{FPR and FNR of GRAMM recognition of recombination events in the simulations with recombination. Note that here the recombination rates are at the family level and not at the individual level, and simulations differ in the number of children.}
\label{table: FPR and FNR of GRAMM}
\end{table}

\FloatBarrier

\subsection{Application to sample families from Australia and\\ Israel}
We further tested the applicability of GRAMM to Australian and Israeli families, and estimated the recombination rate. After filtering families that do not  include at least two family members, including one child, we obtained 996 Australian and 152 Israeli families.

Both Australian and Israeli datasets consist of a combination of serology and DNA-based HLA typing data.
We analyzed the Australian and the Israeli families - molecular and serology typing  (Table~\ref{table: Australian and Israel results}). A significant fraction of the families had inadequate HLA typing information preventing inference of the parents. 
Such families typically contain only children whose alleles do not provide enough data for mapping the parents haplotype (if a family includes children only, GRAMM requires that at least in one locus, these children have 4 different unique alleles (see section~\ref{GRAMM explanation})). The families with missing data constitute about 15\%-25\% of the data (depending on the dataset). Another reason for rejecting a family is two many alleles (more than 4 different alleles in one locus), which can be caused by mutation events, or inclusion of non-nuclear family members: 5\% of the data, or inconsistency of alleles between family member (18\%).  We estimated the recombination rate in the population according to these results, as defined by the fraction of imputed families (families with enough information)  that can only be explained with recombination, with a maximal fraction of 10\%-20\%  recombinations per family, or 1-4 \% recombination rate per child. At this stage, GRAMM does not predict the loci between which the recombination occurred. 

The runtime of GRAMM on our web server is around 4-8 seconds per family (for a single family), and around 1-2 seconds per family, when many families are imputed. There is no practical limit to the number of families that can be estimated. GRAMM is available as a stand alone Python code at \url{https://github.com/zuriyaAnsbacher/GRAMM}  or as a server at \url{https://gramm.math.biu.ac.il/}
\begin{table}[]
\begin{tabular}{|ll|l|l|l|l|}
\hline
\multicolumn{2}{|l|}{} & \textbf{\begin{tabular}[c]{@{}l@{}}Australia\\ (molecular)\end{tabular}} & \textbf{\begin{tabular}[c]{@{}l@{}}Australia\\ (serology)\end{tabular}} & \textbf{\begin{tabular}[c]{@{}l@{}}Israel\\ (molecular)\end{tabular}} & \textbf{\begin{tabular}[c]{@{}l@{}}Israel\\ (serology)\end{tabular}} \\ \hline
\multicolumn{2}{|l|}{\textbf{Success}} & 458 & 142 & 14 & 47 \\ \hline
\multicolumn{1}{|l|}{\multirow{3}{*}{\textbf{Rejection}}} & Missing data & 103 & 35 & 17 & 25 \\ \cline{2-6} 
\multicolumn{1}{|l|}{} & \begin{tabular}[c]{@{}l@{}}Too many \\ values in \\ an allele\end{tabular} & 35 & 18 & 1 & 39 \\ \cline{2-6} 
\multicolumn{1}{|l|}{} & \begin{tabular}[c]{@{}l@{}}Contradiction \\ between family \\ members data\end{tabular} & 122 & 53 & 0 & 9 \\ \Xhline{3\arrayrulewidth}
\multicolumn{1}{|l|}{\multirow{2}{*}{\textbf{\begin{tabular}[c]{@{}l@{}}Success\\ percentage\end{tabular}}}} & \begin{tabular}[c]{@{}l@{}}Considering \\ missing data \\ as a rejection\end{tabular} & 63.8\% & 57.2\% & 43.7\% & 39.1\% \\ \cline{2-6} 
\multicolumn{1}{|l|}{} & \begin{tabular}[c]{@{}l@{}}Ignoring \\ missing data\end{tabular} & 74.5\% & 66.6\% & 93.3\% & 49.4\% \\ \hline
\end{tabular}
\caption{\textbf{Australian and Israeli families rejection and imputation accuracy}. Rejection reasons are separated into 3 groups: (1) Missing data: Family with no parents and none of the  loci contain at least 4 unique alleles (see section~\ref{GRAMM explanation}); (2) Too many alleles in a single locus (more than 4); (3) Contradiction between family members data. Rejection can happen at the first stage of GRAMM, when GRAMM compares children and parents' data to produce a list of candidate haplotypes, or at the final stage, after GRIMM, when the imputed parents' haplotypes are compared to each child's typing.}
\label{table: Australian and Israel results}
\end{table}

\FloatBarrier

\section{Discussion}

Inference of haplotype phasing of HLA in individuals and families is often limited by typing ambiguity. The combination of phase and allele ambiguity require a parallel resolution of the two forms of ambiguity. Historically, the imputation of HLA haplotypes in families was performed by first resolving the allelic ambiguity in each person and then solving the phase ambiguity. 

We have here presented an extension of our previous HLA imputation algorithm, GRIMM \cite{maiers2019grimm}. GRIMM  is based on opening all possible phases, and then resolving the allele ambiguity within each phase. We extended the individual version of GRIMM to the family version - GRAMM, by first imputing only the (presumed) parents, and subsetting the list of possible phases  based on consistency with children's possible haplotypes assuming inheritance of complete haplotypes (in all children up to at most one recombination event).

We have shown that GRAMM is more accurate than GRIMM in phase resolution, and sometimes also in resolving allelic ambiguity in both simulated family data, pairs of mother and their cord blood, and in works on families from Israel and Australia. We have further shown that GRAMM can be used to estimate the recombination rates, which was observed in families to be 1-4 \% per person per generation. This is higher than previously estimated at most 1 \%\cite{blancher2012use}. 

GRAMM contains three main components: the first is a phase locking/reduction scheme, based on the inheritance of full haplotypes. The next step is followed by the imputation of the reconstructed parents typing. The last stage is a rejection of  solutions not consistent with all family members (up to one member that can be rejected based on possible recombination events). The GRAMM algorithm approach is in contrast with all existing approaches that either ignore ambiguity, or first resolve it at the single person level, and only then estimate inheritance.

GRAMM can be used not only for large-scale analysis of family HLA typing for research, but also for practical usage in clinical histocompatibility laboratories and transplant centers. Moreover, it can impute alleles not typed in any family member. Family HLA imputation is often required in related donors transplant. The imputation of all family members, which is currently typically performed manually, can be less time-consuming and more accurate using GRAMM.
GRAMM has two main limitations. A) It does not handle complex (non-nuclear families). B) It does not handle homozygote parents, when the input contains no parents typing. However, in such cases family imputation is trivial. 
\section{Methods}
\subsection{Data}
\subsubsection{Cord blood data}
We analyzed the HLA from 16,404 cord blood units with accompanying maternal HLA typing described previously \cite{magalon2015banking}.

\begin{table}[]
\begin{tabular}{|c|ccccc|ccccccccc|}
\hline
\multirow{3}{*}{} & \multicolumn{5}{c|}{\textbf{Parents number}} & \multicolumn{9}{c|}{\textbf{Children number}} \\ \cline{2-15} 
 & \multicolumn{3}{c|}{\textbf{Distribution}} & \multicolumn{1}{c|}{\multirow{2}{*}{\textbf{Avg}}} & \multirow{2}{*}{\textbf{Med}} & \multicolumn{7}{c|}{\textbf{Distribution}} & \multicolumn{1}{c|}{\multirow{2}{*}{\textbf{Avg}}} & \multirow{2}{*}{\textbf{Med}} \\ \cline{2-4} \cline{7-13}
 & \multicolumn{1}{c|}{0} & \multicolumn{1}{c|}{1} & \multicolumn{1}{c|}{2} & \multicolumn{1}{c|}{} &  & \multicolumn{1}{c|}{0} & \multicolumn{1}{c|}{1} & \multicolumn{1}{c|}{2} & \multicolumn{1}{c|}{3} & \multicolumn{1}{c|}{4} & \multicolumn{1}{c|}{5} & \multicolumn{1}{c|}{6+} & \multicolumn{1}{c|}{} &  \\ \hline
\textbf{Aus} & \multicolumn{1}{c|}{434} & \multicolumn{1}{c|}{210} & \multicolumn{1}{c|}{490} & \multicolumn{1}{c|}{1.05} & 1 & \multicolumn{1}{c|}{6} & \multicolumn{1}{c|}{225} & \multicolumn{1}{c|}{378} & \multicolumn{1}{c|}{280} & \multicolumn{1}{c|}{146} & \multicolumn{1}{c|}{53} & \multicolumn{1}{c|}{54} & \multicolumn{1}{c|}{2.68} & 2 \\ \hline
\textbf{Isr} & \multicolumn{1}{c|}{164} & \multicolumn{1}{c|}{0} & \multicolumn{1}{c|}{0} & \multicolumn{1}{c|}{0} & 0 & \multicolumn{1}{c|}{0} & \multicolumn{1}{c|}{0} & \multicolumn{1}{c|}{68} & \multicolumn{1}{c|}{25} & \multicolumn{1}{c|}{19} & \multicolumn{1}{c|}{15} & \multicolumn{1}{c|}{37} & \multicolumn{1}{c|}{3.88} & 3 \\ \hline
\end{tabular}
\caption{\textbf{Australian and Israeli data distribution.} The distribution of the number parents and children is presented, including their average and median. Note that we include all the data (even families without children or less than 2 individuals); therefore, the sum of the numbers is higher than in Table ~\ref{table: Australian and Israel results}.}
\label{table: Australian and Israeli data distribution}
\end{table}

\subsubsection{Australian and Israeli families}
We tested 996 real families from Australia and 152 from Israel, see Table ~\ref{table: Australian and Israeli data distribution} for data distribution. For each family member, typing was reported at varying levels of resolution. The Israeli families were obtained from Beilinson Hospital in Israel, and the Australian families were obtained from collaborators in the Department of Clinical Immunology, PathWest, Fiona Stanley Hospital, Perth, Australia. 
Both Australian and Israeli  families typed by a mixture of serology and molecular methods (a family is a serology family if there is at least one family member with a serology sample). Therefore, we split the data into four groups: Australian (genetic), Australian (serology), Israel (genetic) and Israel (serology). 

\subsubsection{Simulations of HLA typing of family pedigrees.}\label{simulations explanation}
Simulated HLA typing files were generated by sampling 2 haplotypes for each individual from NMDP full registry haplotype distributions. Each simulation contains 500 families. Each row is a person - parent or children, and a genotype consisting of 2 haplotypes and the source population. Current pedigree simulations are for the US Caucasian (CAU) population only. Simulated HLA typing data for families and simulation code is publicly available on GitHub:  \url{https://github.com/nmdp-bioinformatics/ihiw-pedigree-sim}. The input to 
\subsection{GRIMM}
GRIMM is the unphased and ambiguous genotype of each family member represented in genotype list string (GL string) format \cite{milius2013genotype}, and the output is the list of the 10 most probable haplotype pairs, based on priors of donor/patient ethnicity information and population HLA haplotype frequency distributions for each population.

\subsection{GRAMM Pedigree-based HLA Imputation Algorithm Schematic} \label{GRAMM explanation}

\FloatBarrier
\begin{itemize}
    \item The input is a GL-string of the family members \cite{milius2013genotype}.
    \item GRAMM is only imputing parents, and is initialized with 2 empty ``haplotypes" for each parent. The children haplotypes are computed from the parents.
\item Typing data input \newline
 GRAMM uses two possibilities. Either parents exist in the data of the family (at least one), or not.
 \begin{itemize}
     \item Parents exist - The parents' typing is used for the parents initial filling of possible haplotype combinations: two values of each allele in the genotype are inserted into the two haplotype, except for the first allele, where we insert a single value into each haplotype (because the order between the haplotypes does not matter). In addition, if an allele is a homozygous, we assign it to both alleles. For example, given a parent genotype: A*01+A*02\textsuperscript{$\wedge$}B*03+
    B*04\textsuperscript{$\wedge$}C*05+C*06\textsuperscript{$\wedge$}DRB1*07
     +DRB1*07\textsuperscript{$\wedge$}DQB1*08+DQB1*09, 
     the insertion will be: Haplotype 1: A: 01, B: 03/04, C:05/06, DRB1: 07, DQB1: 08/09, and  Haplotype 2: A: 02, B: 03/04, C: 05/06, DRB1: 07, DQB1: 08/09.
\item If there are no parents in the data, we look for an HLA locus that has four unique alleles among all the children. If such a locus exists, we insert the alleles into the parents haplotypes, assuming inheritance (else, the process is stopped). Note that this is the main weakness of GRAMM - It does not handle homozygote parents, when the input contains no parents typing. However, in such cases family imputation is trivial. 
 
For example, given the following children:

child 1: A*01+A*02, child 2: A*02+A*03, child 3: A*03+A*04

The insertion will be:
\begin{table}[h!]
\centering
\begin{tabular}{l|cc}
           & Parent 1 & Parent 2 \\ \hline
Haplotype 1 & 01                & 02                \\
Haplotype 2 & 03                & 04               
\end{tabular}
\end{table}
 \end{itemize} 

\item Matching and adding loci. Once we are guaranteed that both parents' haplotypes contain at least one locus with at least an allele in each haplotype,  we can associate between each child and the 2 haplotypes inherited from the parents (one from each one of them). For instance, in the example above, child `2' inherited haplotype `2' from first parent and haplotype `1' from the second parent.

Once we have the information about the 2 (partial) haplotypes each child inherited from their parents, we loop over the remaining loci of each child, and compare them to the alleles in the haplotypes they inherited from the parents, and reduce the phasing options to the alleles in the parents' haplotypes. 
 
For example, given the following allele in parent haplotype: B*05/06 and a child genotype that contains: B*05+B*08, we can determine that the allele in the parental haplotype is B*05.

At this stage, if there is a contradiction between the data of family members, the family is rejected. The contradiction could be caused by recombination, incorrect typing or a mutation. If a single recombination is observed, the appropriate child is ignored. In all cases, GRAMM reports the cause of failure.
\item Imputation by GRIMM. \newline
The  parents' haplotypes can be ambiguous for two reasons:
\begin{enumerate}
    \item Incomplete HLA genotyping data. 
    \item The family can have unresolved typing ambiguity even after pedigree inference.
\end{enumerate}
An improved version of GRIMM \cite{maiers2019grimm} is used to handle both problems through imputation.

In the GRIMM imputation, each genotype should include two alleles for each typed locus (that may be the same).  GRIMM can not handle a case of one haplotype with a typing, and the second haplotype without. Often, families consist of one parent and one child only. In such cases, the second parent has an haplotype with no data. To avoid removing the haplotype with the data, we duplicate the single known haplotype and insert to GRIMM for imputation. After the imputation, we keep the only the real imputed haplotype and the artificial haplotype is removed.
\item Following imputation, we check all the pairs of haplotypes consistent with the parents typing with a high enough probability ($p<1.e-15$) in the imputation output and test their consistency with the children typing. Inconsistent parents' haplotypes pairs are rejected. If all the possible parents haplotype pairs are inconsistent, the family is rejected. However, this happens very rarely.

\end{itemize}

\subsection{GRIMM -phase locking, single population }
GRIMM \cite{maiers2019grimm} uses as input a GL String and the ethnicity and returns the consistent haplotype pairs and probabilities conditional on reference population haplotype frequencies. GRIMM first solves the phase ambiguity according to the constraints from the family's information, such that alleles that are known to be in the same haplotype must be in the same phase and then within each phase resolves the allele ambiguity to produce a candidate (partial) high-resolution haplotype pair. The probability of this haplotype pair is calculated as the product of the probabilities of each haplotype in the reference population haplotype frequencies. These probabilities are pre-loaded into nodes for each HLA allele and haplotype for each sub-population. To obtain the probability of each possible haplotype, a query to a graph is performed. Then a Cartesian product is performed on all legitimate haplotype pairs to produce all probable enough haplotype pairs \cite{israeli2021hla}.

\subsection{Website input and output format} \label{web explanation}
\begin{figure}
\centering

    \begin{subfigure}{0.9\columnwidth}
    \centering
    \includegraphics[width=1\textwidth]{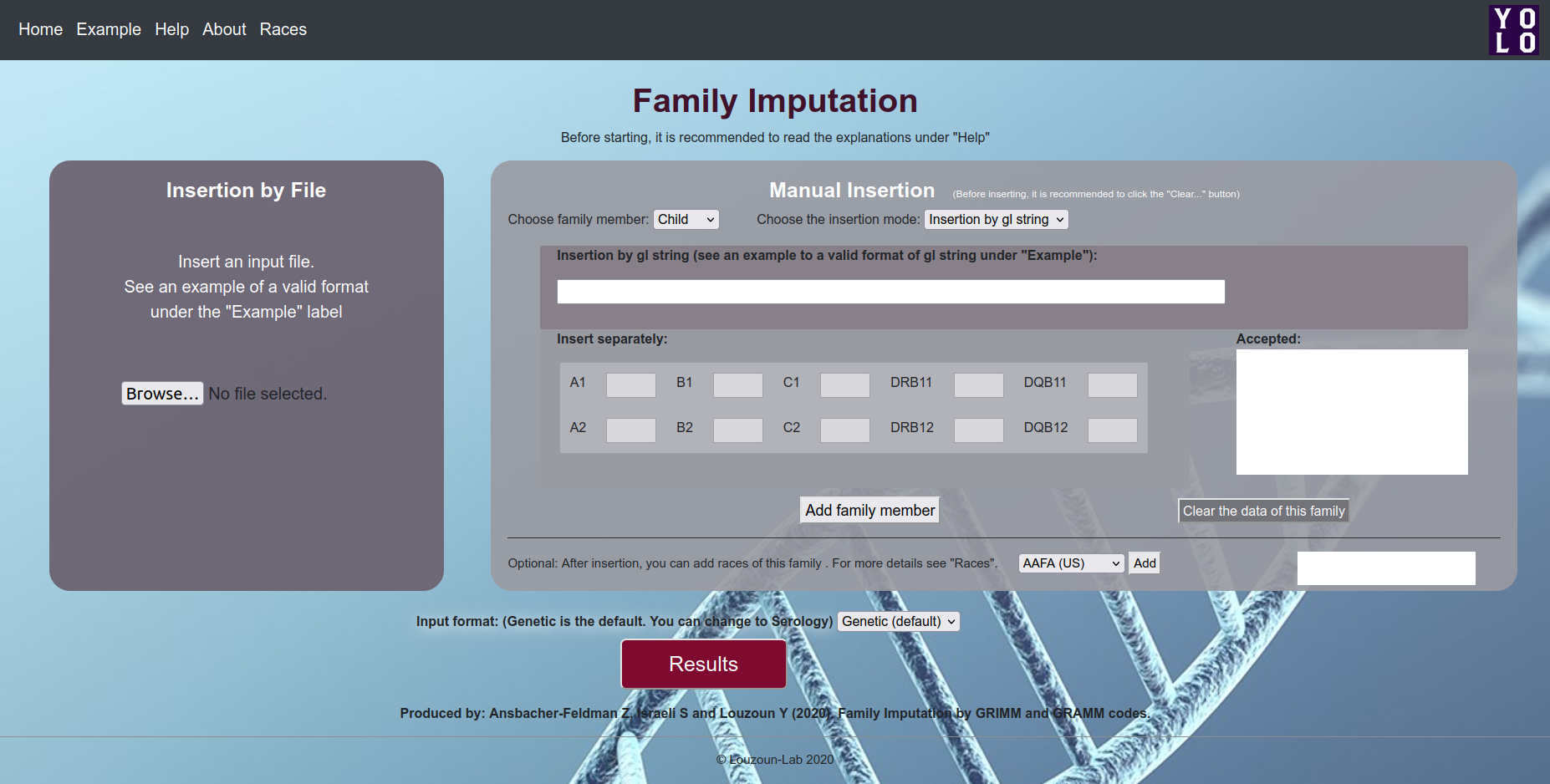}
    \caption{GRAMM website}
    \end{subfigure}
    
    \begin{subfigure}{0.9\columnwidth}
    \centering
    \includegraphics[width=1\textwidth]{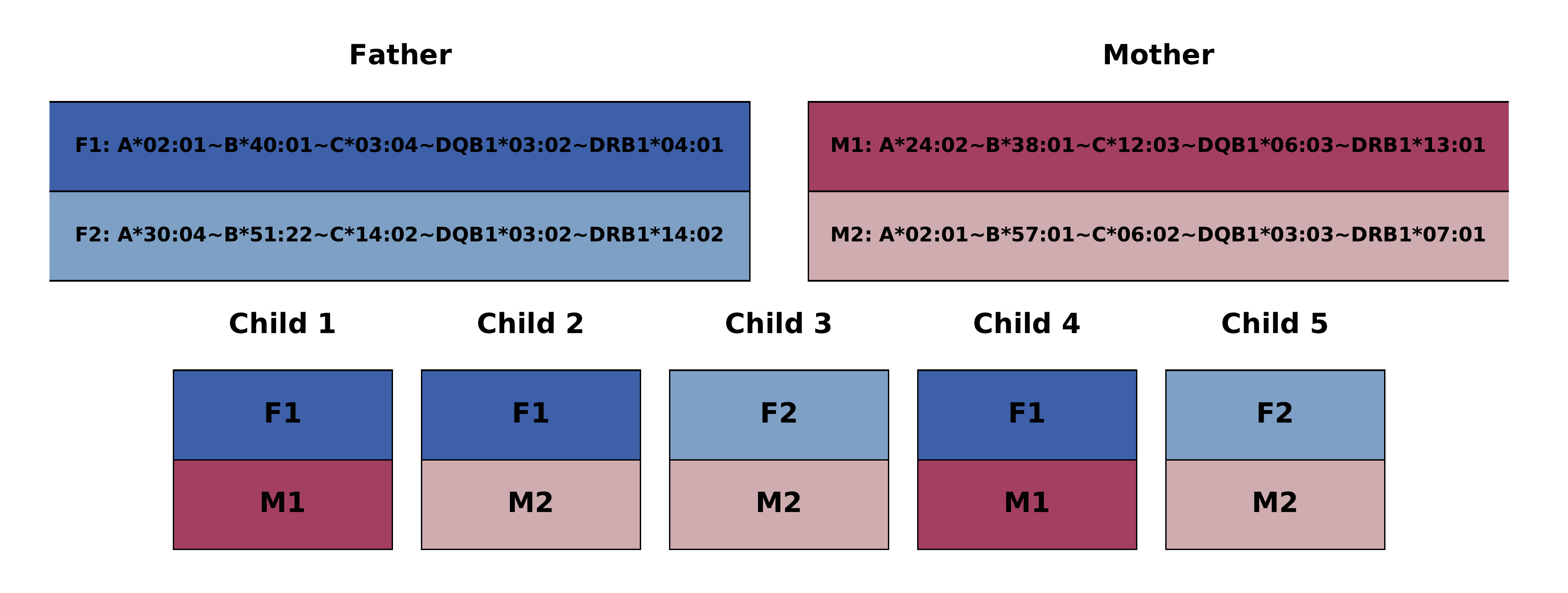}
    \caption{Pedigree visualization}
    \end{subfigure}

\caption{\textbf{GRAMM website:} (a) A screenshot from the internet page. (b) An example of a pedigree visualization that the user receives from the GRAMM website. F1 and M1 stand for the first paternal and maternal haplotypes.}
\label{fig: GRAMM website screenshot and pedigree visualization}
\end{figure}

\begin{itemize}
    \item \textbf{Input format:} single or several families, with HLA typing for each family member. The input can be provided as a CSV file or by manual insertion into a web form. Example input files are available in the server.
    \item \textbf{Output format:} after processing, family imputation output can be downloaded as a ZIP archive, which contains the following files: an imputation output file (with the most likely parents' haplotypes of the inserted families); an errors file (listing the families that did not pass the process successfully, and the failure reason); pedigree visualizations which present the haplotypes of each family members visually (PNG image and PDF formats) (see Figure \ref{fig: GRAMM website screenshot and pedigree visualization}); and in case of manual insertion - an additional file, which includes the input data that the user submitted, for easy adjustment and re-submission, if needed.
\end{itemize}

\bibliographystyle{plain}
\bibliography{references}

\pagebreak
\begin{center}
\textbf{\huge Supplemental Materials}
\end{center}

\setcounter{equation}{0}
\setcounter{figure}{0}
\setcounter{table}{0}
\setcounter{page}{1}
\makeatletter
\renewcommand{\theequation}{S\arabic{equation}}
\renewcommand{\thefigure}{S\arabic{figure}}
\renewcommand{\thetable}{S\arabic{table}}

\begin{figure}[htp]
    \centering
    \includegraphics[width=0.59\linewidth]{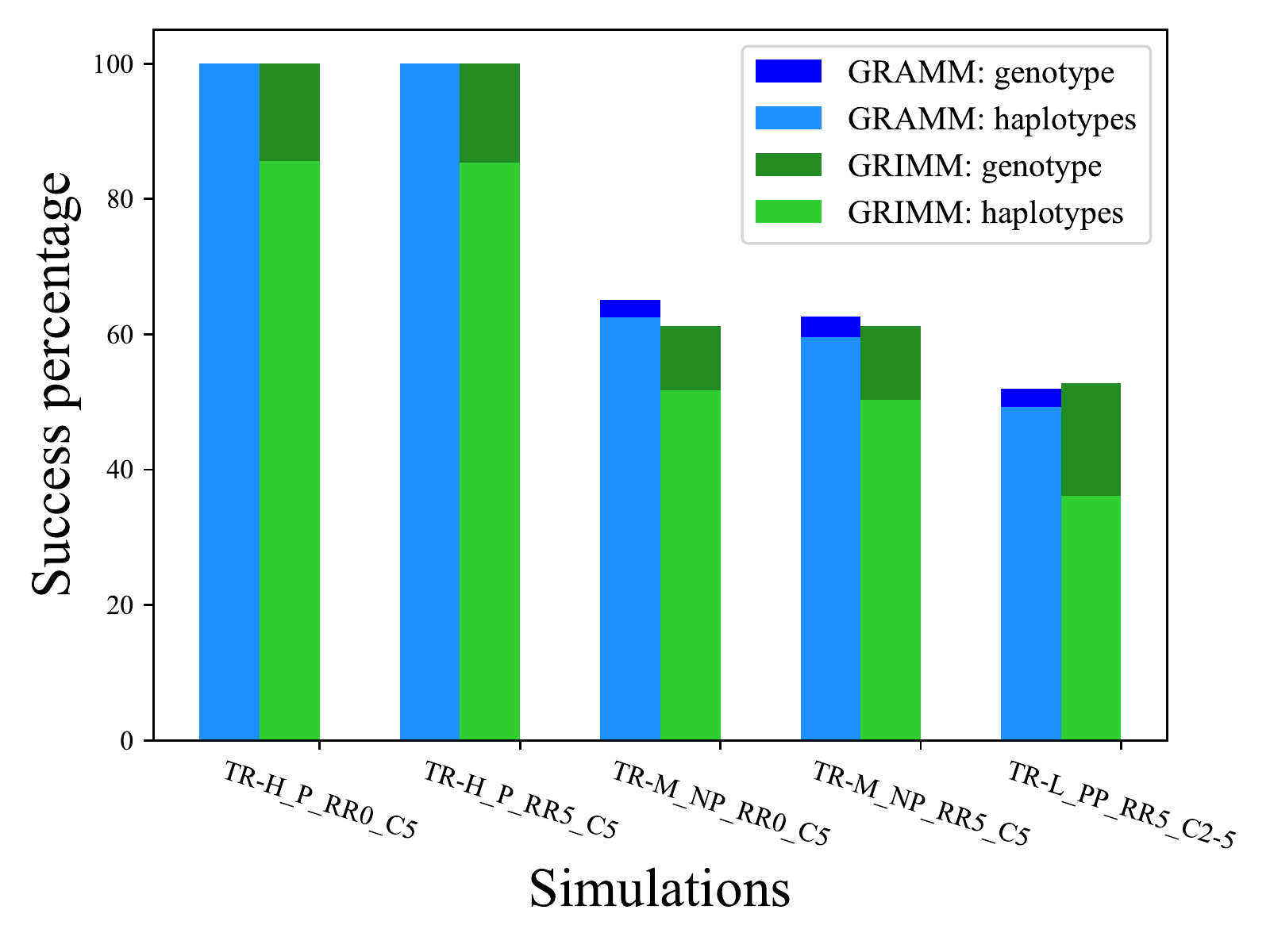}
    \caption{\textbf{Simulations set 2 - Comparison between the accuracy, measured as the fraction of properly imputed individuals haplotypes/genotypes  of GRAMM (blue) and GRIMM (green).} Deep blue and green represent accuracy in the unphased genotypes, and light blue green are for precision in the phased genotypes. The difference between deep and light are errors in phasing.}
    \label{fig: GRAMM VS GRIMM Success - set 2}
\end{figure}

\begin{table}[htp]
\begin{tabular}{cllccc}
\hline
\begin{tabular}[c]{@{}c@{}}Simulation\\ ID\end{tabular} &
  \multicolumn{1}{c}{\begin{tabular}[c]{@{}c@{}}Simulation \\ name\end{tabular}} &
  \multicolumn{1}{c}{\begin{tabular}[c]{@{}c@{}}Typing \\ resolution\end{tabular}} &
  \begin{tabular}[c]{@{}c@{}}Parents\\ number\end{tabular} &
  \begin{tabular}[c]{@{}c@{}}Recombination\\ rate\end{tabular} &
  \begin{tabular}[c]{@{}c@{}}Children\\ number\end{tabular} \\ \hline
1  & TR-H\_P\_RR0\_C5     & High   & 2   & 0  & 5   \\ \hline
2  & TR-H\_P\_RR5\_C5     & High   & 2   & 5  & 5   \\ \hline
3  & TR-M\_NP\_RR0\_C5     & Medium & 0   & 0  & 5   \\ \hline
4  & TR-M\_NP\_RR5\_C5    & Medium & 0   & 5  & 5   \\ \hline
5  & TR-L\_PP\_RR5\_C2-5    & Low & 0-2   & 5  & 2-5   \\ \hline
\end{tabular}
\caption{\textbf{Simulations details - set 2.} Simulations are notated by 4 fields: (1) \textbf{TR}- typing resolution (High = no ambiguity + fully typing (A,B,C,DRB1,DQB1); Medium = ambiguity + fully typing; Low = ambiguity + some missing C/DQB1), (2) \textbf{P/NP/PP}- Number of parents (P =  2 parents; NP =  no parents; PP = partial parents (0-2)), (3) \textbf{RR}- recombination rate (0/1/2/5/10 \%), (4) \textbf{C}- children number (5/2-5)}
\label{table: Simulation details - set 2}
\end{table}
\FloatBarrier

\begin{figure}[htp]
\centering
    \begin{subfigure}{0.45\columnwidth}
    \centering
    \includegraphics[width=\textwidth]{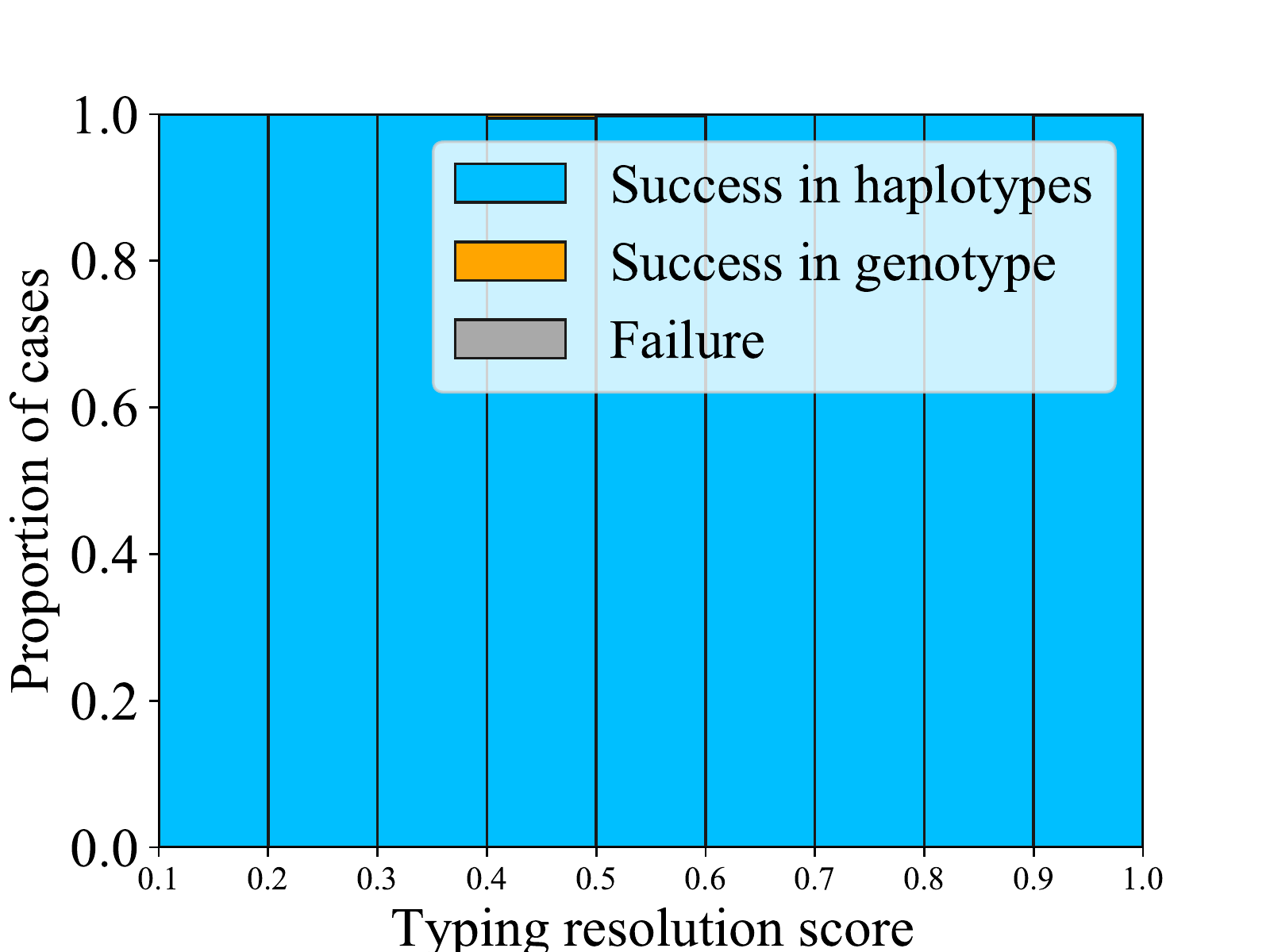}
    \caption{simulation 1: TR-H\_P\_RR0\_C3}
    \label{fig:time1}
    \end{subfigure}
    \begin{subfigure}{0.45\columnwidth}
    \centering
    \includegraphics[width=\textwidth]{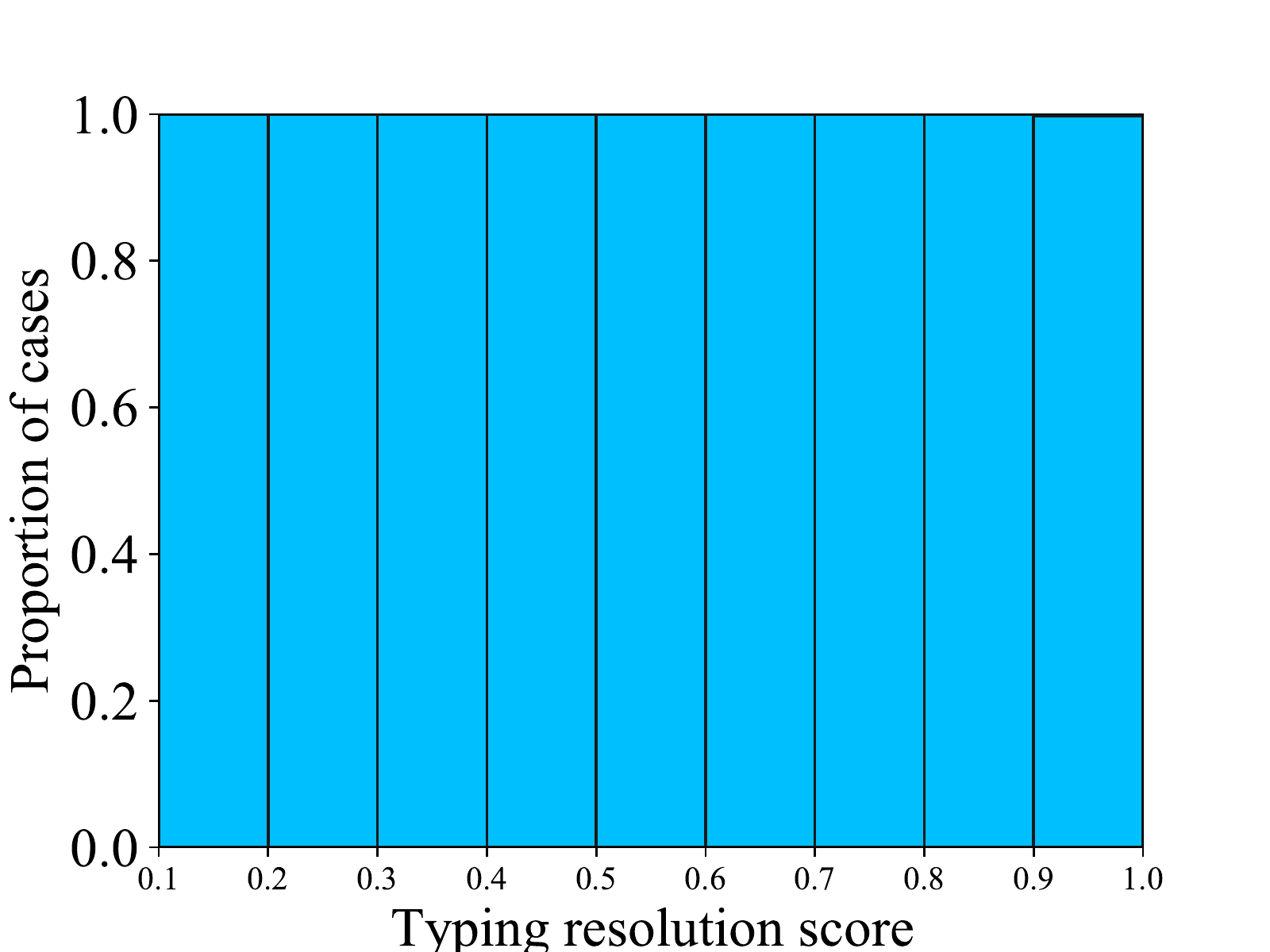}
    \caption{simulation 2: TR-H\_P\_RR5\_C3}
    \label{fig:time2}
    \end{subfigure}

    \begin{subfigure}{0.45\columnwidth}
    \centering
    \includegraphics[width=\textwidth]{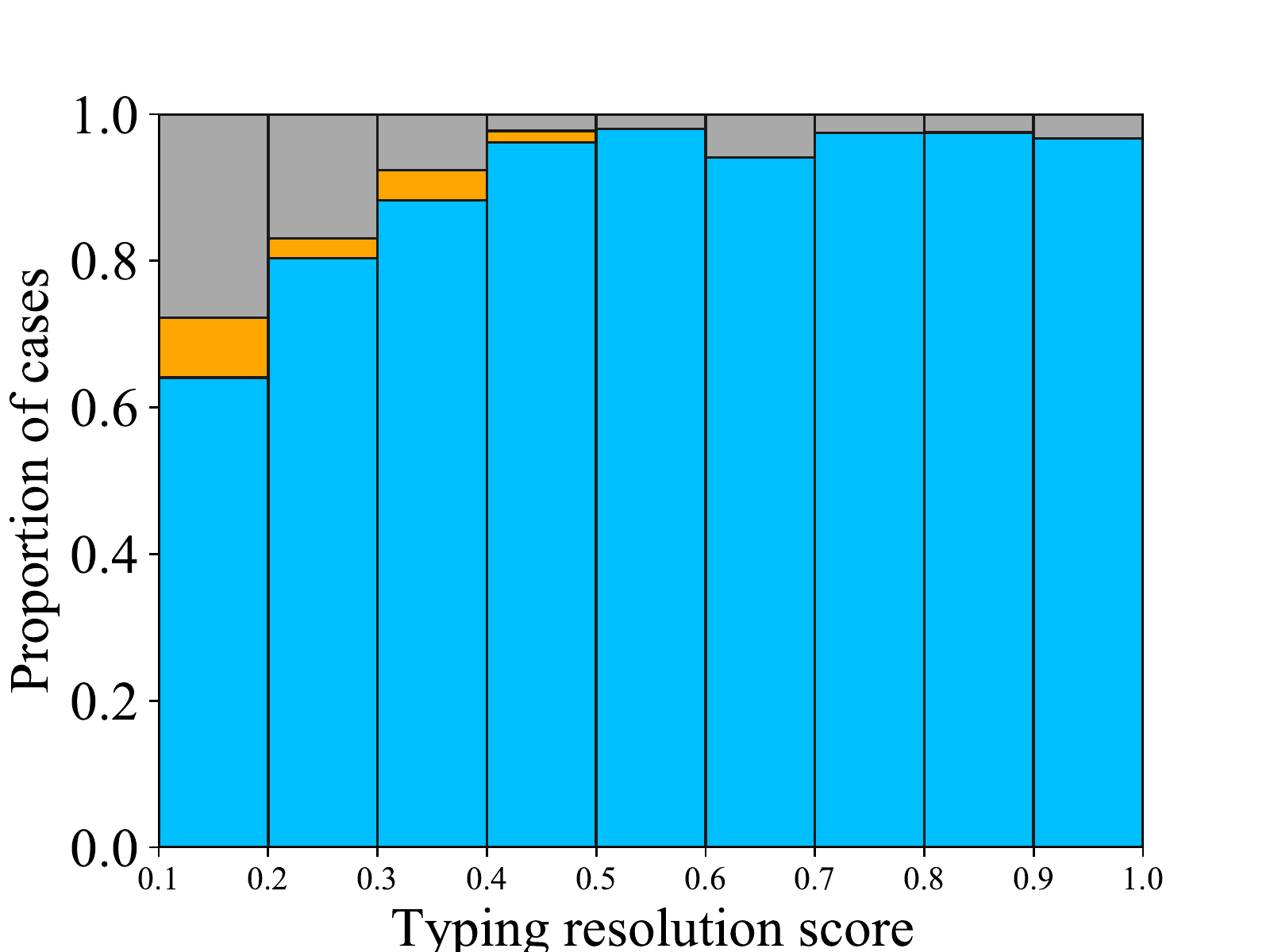}
    \caption{simulation 6: TR-L\_PP\_RR0\_C1-3}
    \label{fig:time1}
    \end{subfigure}
    \begin{subfigure}{0.45\columnwidth}
    \centering
    \includegraphics[width=\textwidth]{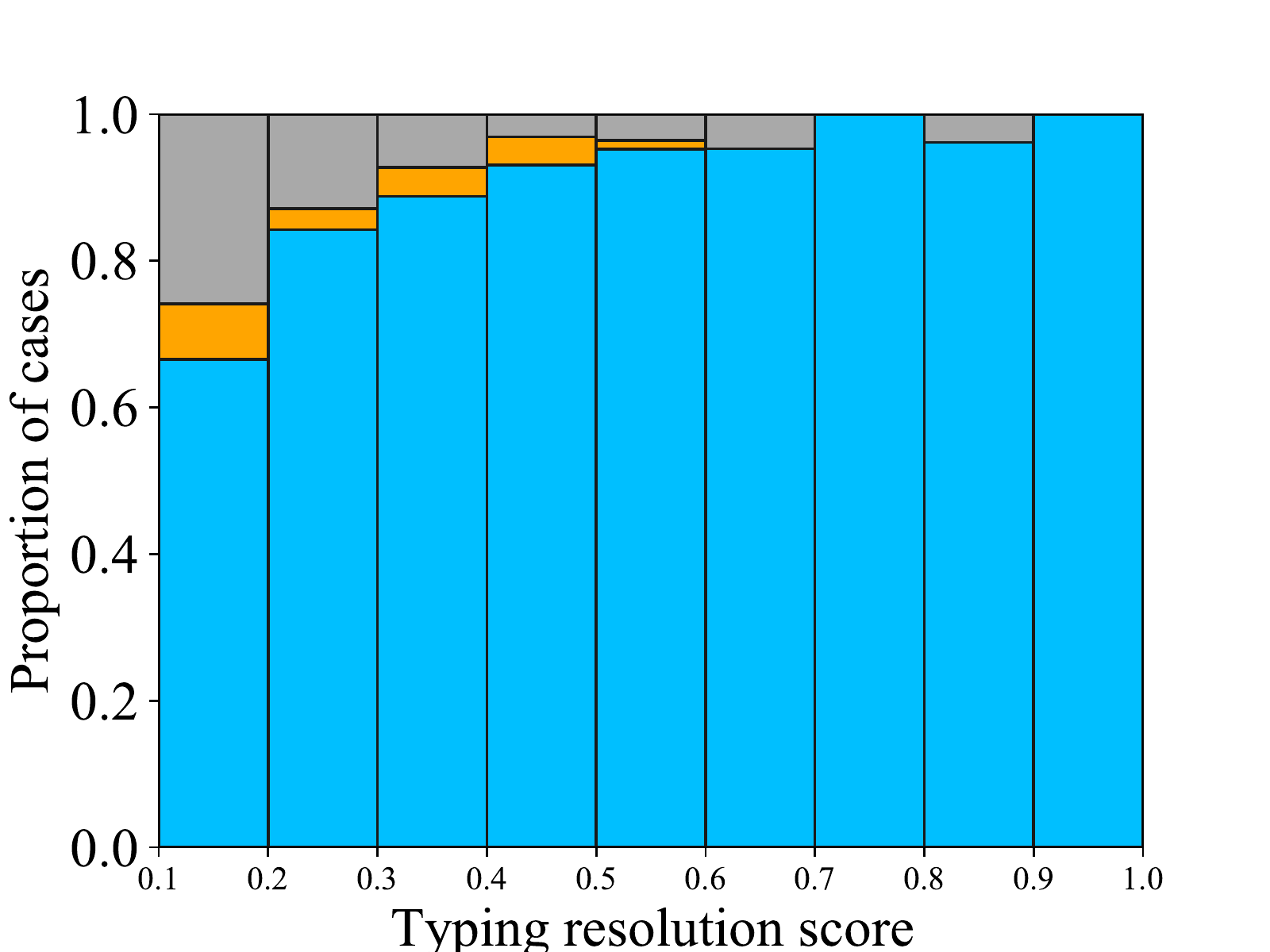}
    \caption{simulation 7: TR-L\_PP\_RR1\_C1-3}
    \label{fig:time2}
    \end{subfigure}
    
    \begin{subfigure}{0.45\columnwidth}
    \centering
    \includegraphics[width=\textwidth]{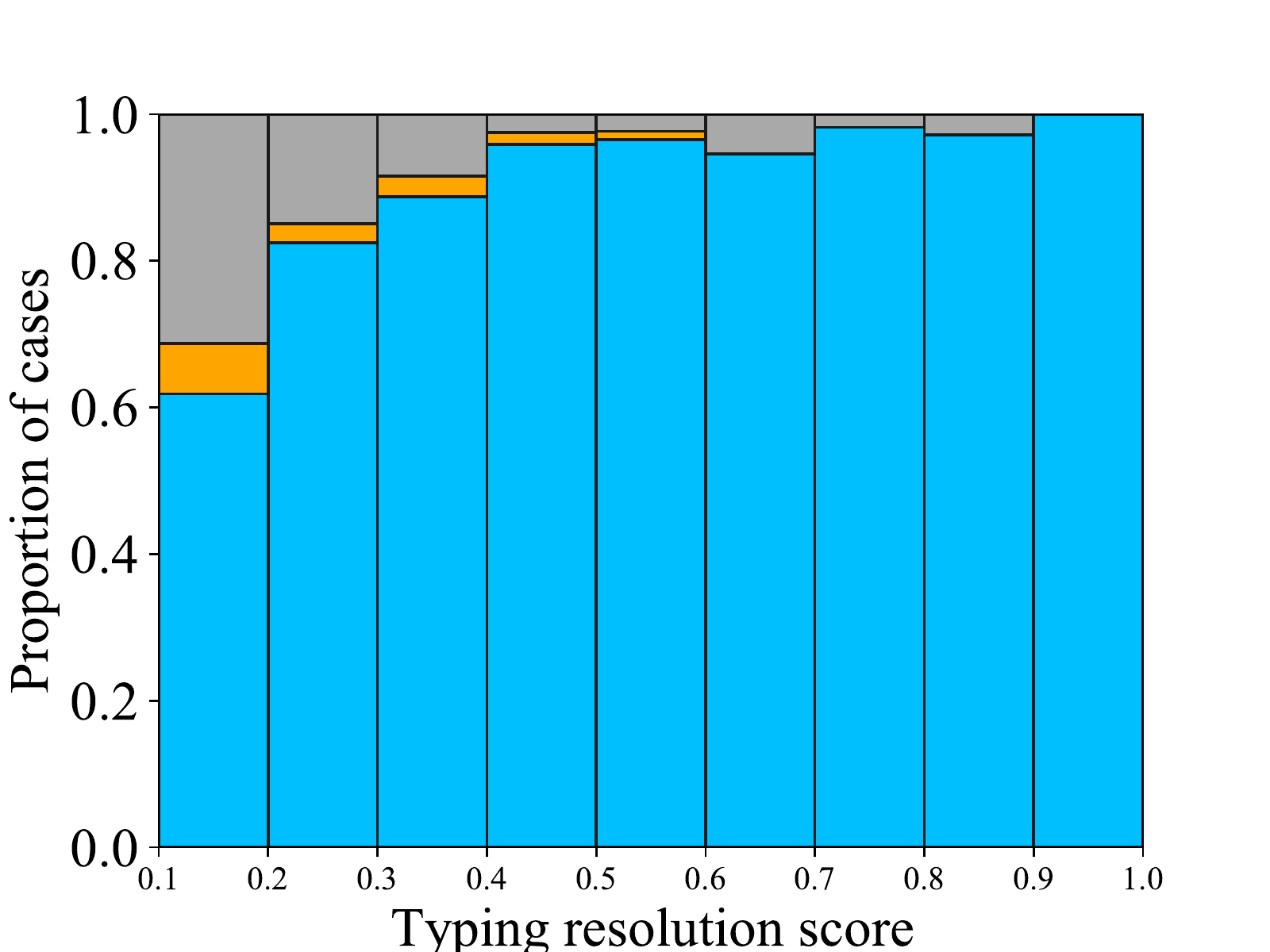}
    \caption{simulation 8: TR-L\_PP\_RR2\_C1-3}
    \label{fig:time1}
    \end{subfigure}
    \begin{subfigure}{0.45\columnwidth}
    \centering
    \includegraphics[width=\textwidth]{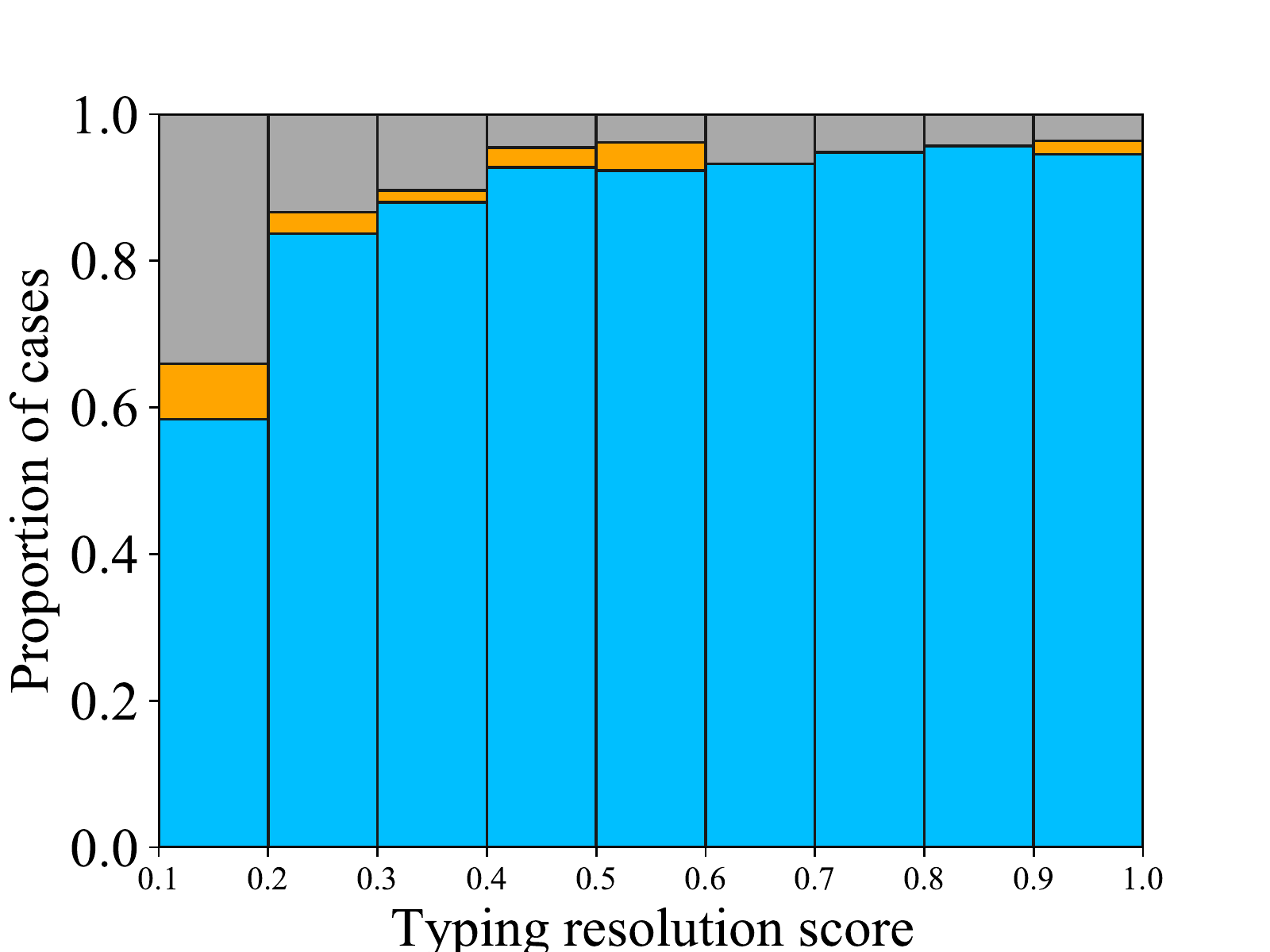}
    \caption{simulation 10: TR-L\_PP\_RR10\_C1-3}
    \label{fig:time2}
    \end{subfigure}
    
    \caption{\textbf{Error rate vs $S=\sum_i P_i^2$.} We computed for each simulated donor $S=\sum_i P_i^2$, and for each set of donors grouped by $S$, we computed the fraction of properly computed genotypes (Blue), proper genotype, but wrong phasing (Orange), and wrong alleles (Gray). The error rate decreases with $TRS$, but increase in the more complex simulations. Presented here are 6 simulations that supplement the 4 presented in the original paper.}
\end{figure}

\begin{figure}[htp]
\centering

    \begin{subfigure}{0.45\columnwidth}
    \centering
    \includegraphics[width=\textwidth]{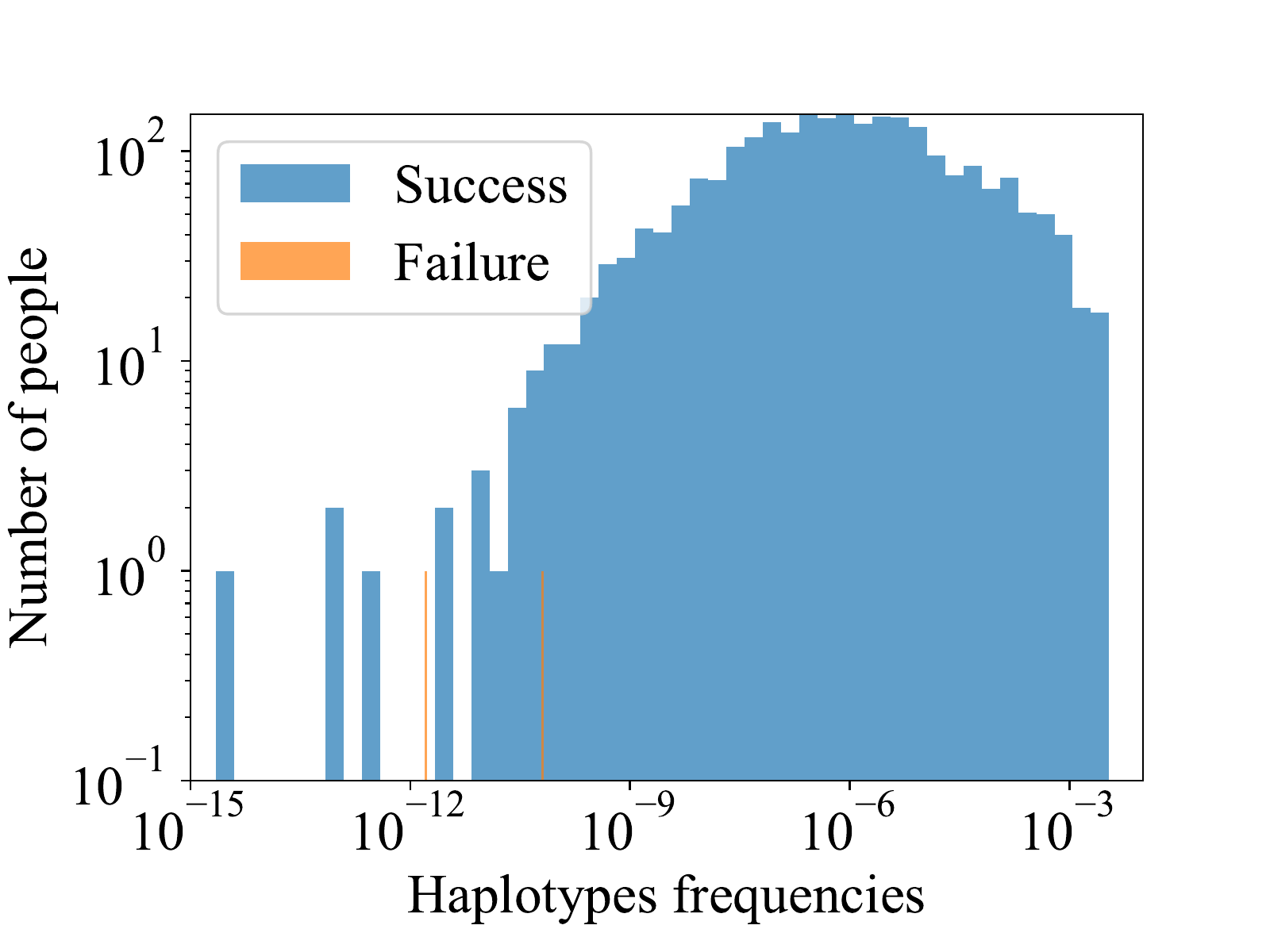}
    \caption{simulation 1: TR-H\_P\_RR0\_C3}
    \label{fig:time1}
    \end{subfigure}
    \begin{subfigure}{0.45\columnwidth}
    \centering
    \includegraphics[width=\textwidth]{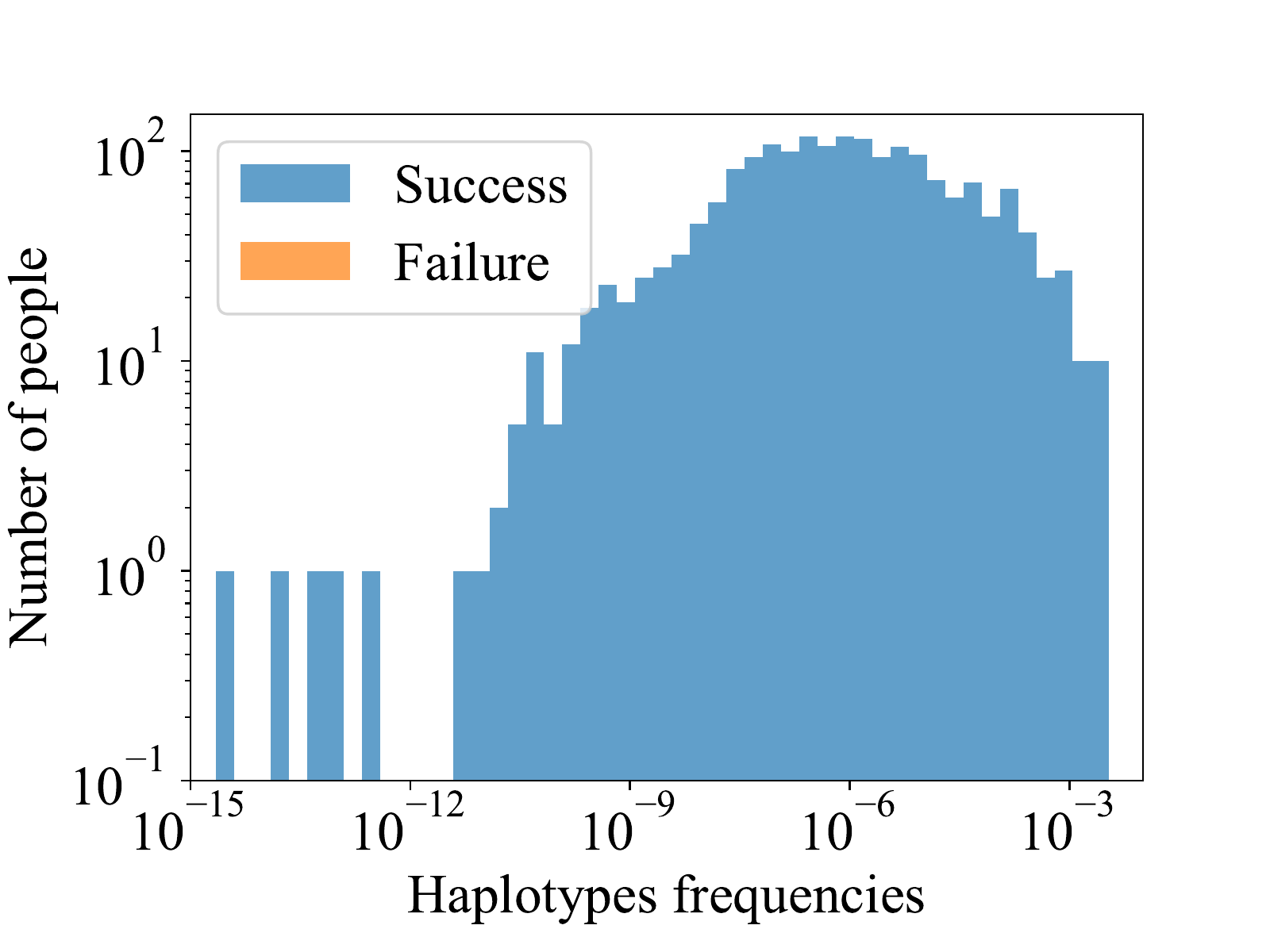}
    \caption{simulation 2: TR-H\_P\_RR5\_C3}
    \label{fig:time2}
    \end{subfigure}

    \begin{subfigure}{0.45\columnwidth}
    \centering
    \includegraphics[width=\textwidth]{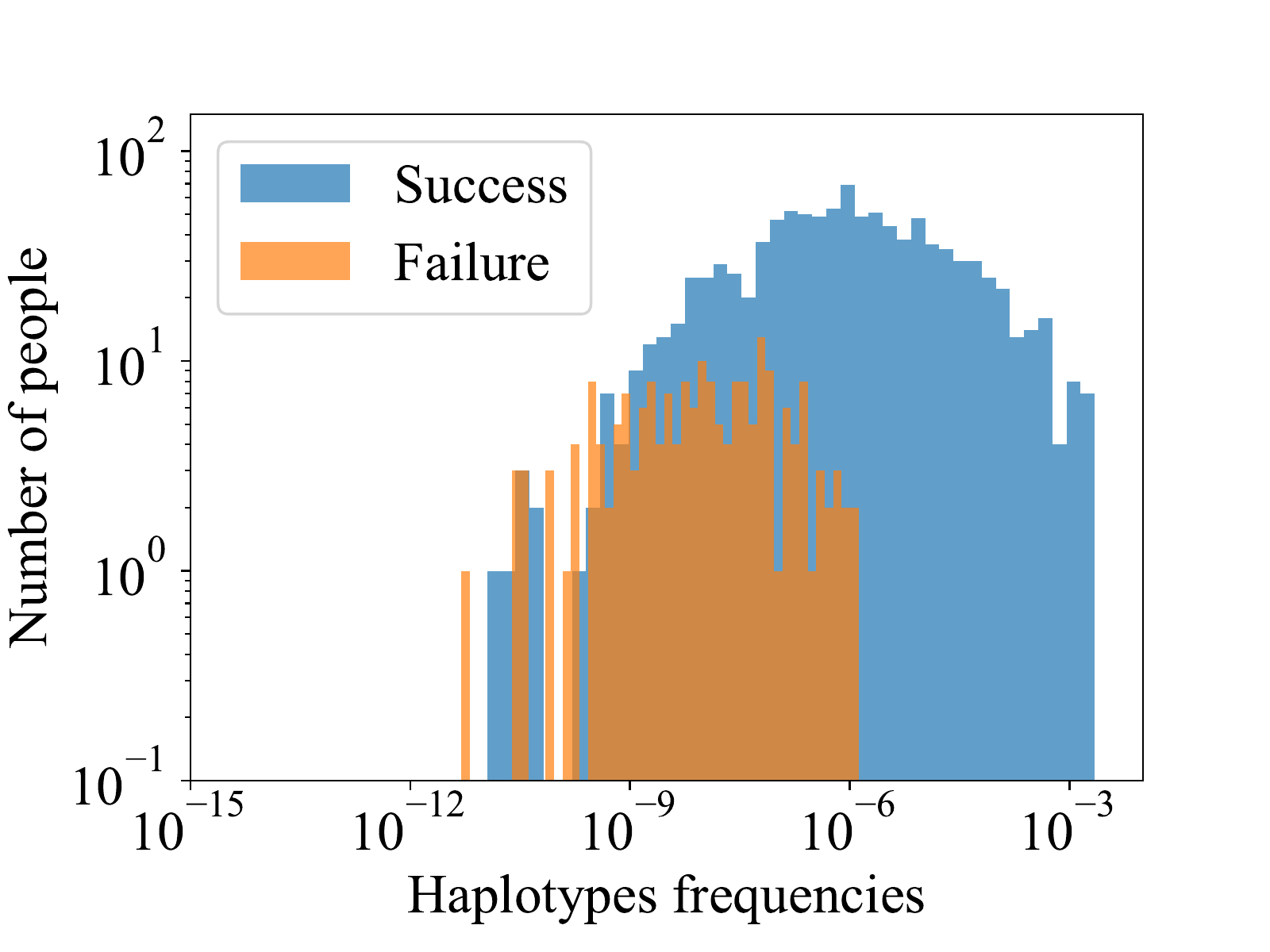}
    \caption{simulation 6: TR-L\_PP\_RR0\_C1-3}
    \label{fig:time1}
    \end{subfigure}
    \begin{subfigure}{0.45\columnwidth}
    \centering
    \includegraphics[width=\textwidth]{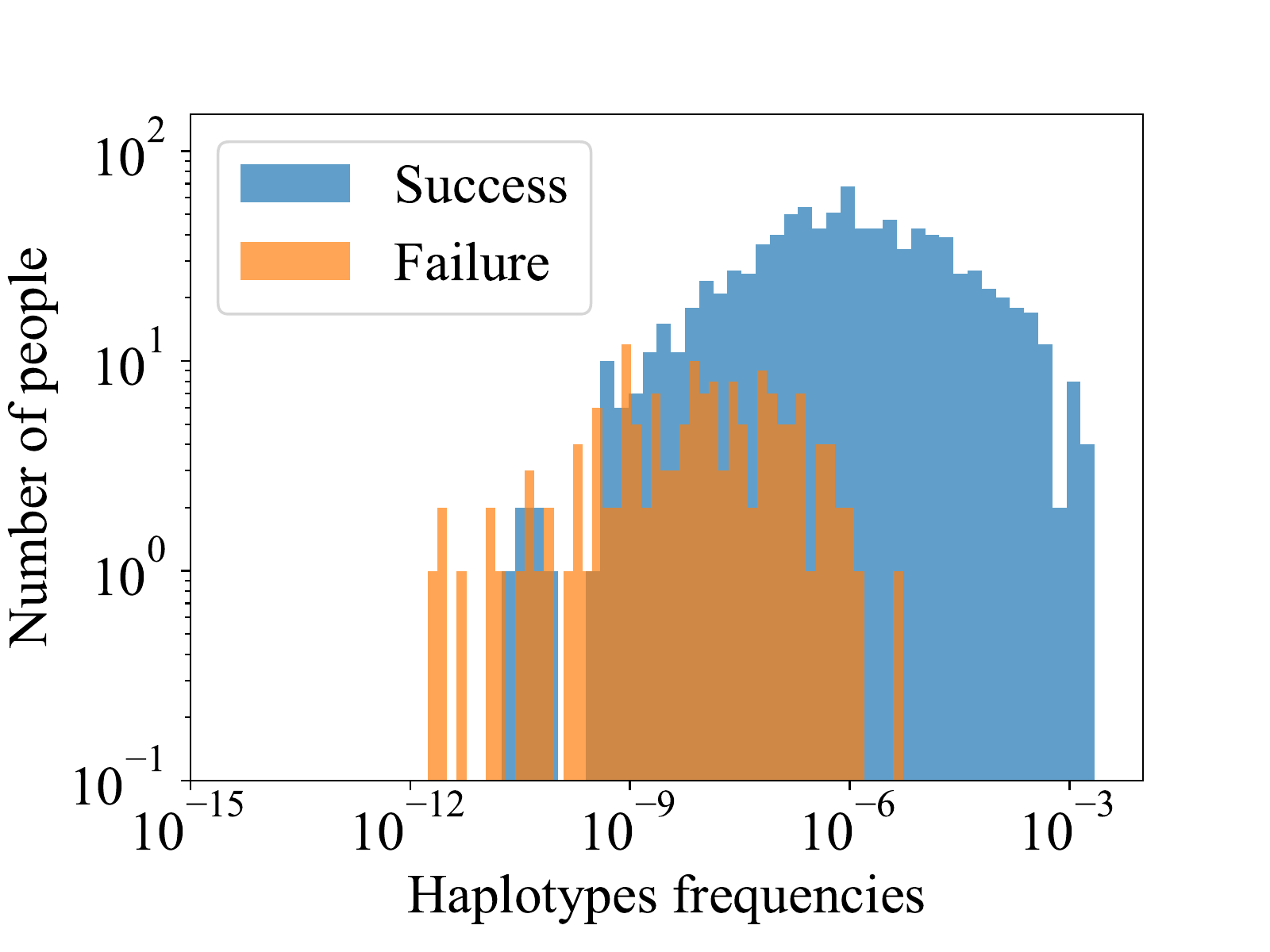}
    \caption{simulation 7: TR-L\_PP\_RR1\_C1-3}
    \label{fig:time2}
    \end{subfigure}
    
    \begin{subfigure}{0.45\columnwidth}
    \centering
    \includegraphics[width=\textwidth]{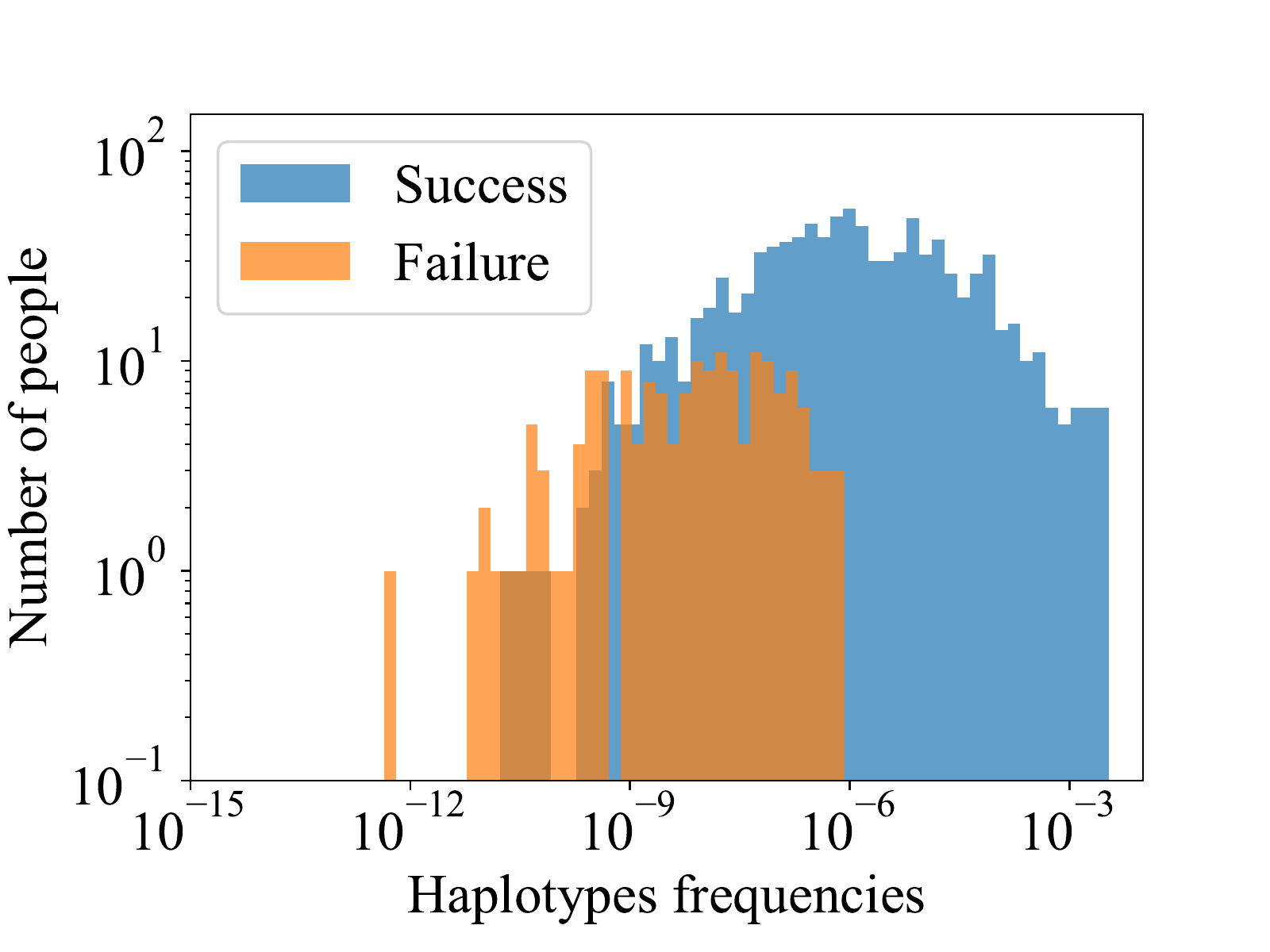}
    \caption{simulation 8: TR-L\_PP\_RR2\_C1-3}
    \label{fig:time1}
    \end{subfigure}
    \begin{subfigure}{0.45\columnwidth}
    \centering
    \includegraphics[width=\textwidth]{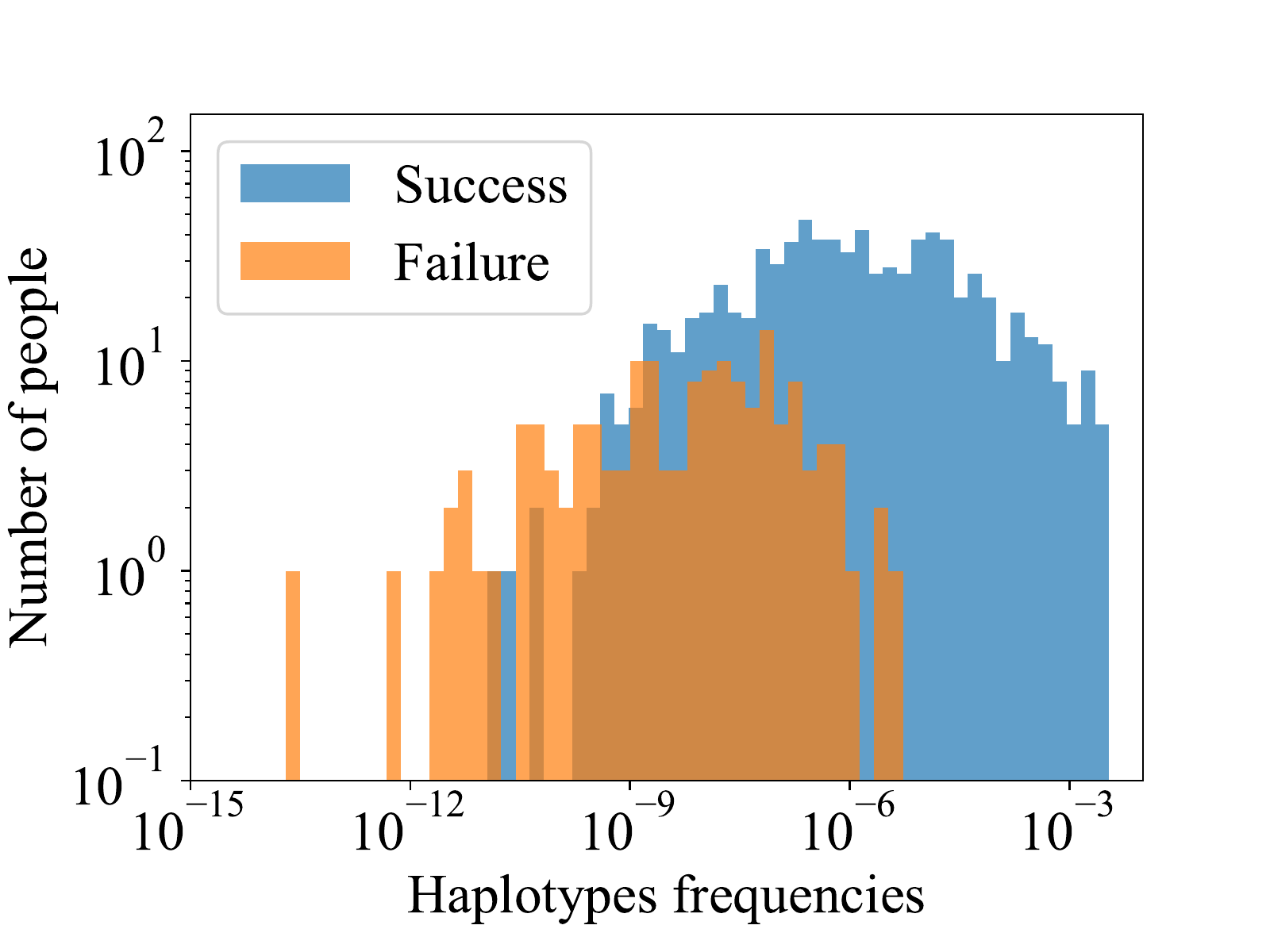}
    \caption{simulation 10: TR-L\_PP\_RR10\_C1-3}
    \label{fig:time2}
    \end{subfigure}
    
    \caption{Haplotype frequencies of properly imputed genotypes (blue bars) and erroneously imputed genotypes (orange bars). Presented here are 6 simulations that supplement the 4 presented in the original paper.}
\end{figure}

\end{document}